\documentclass[10pt]{iopart}

\usepackage[T1]{fontenc}

\usepackage[usenames,dvipsnames]{color}

\definecolor{webgreen}{rgb}{0,0.75,0}
\definecolor{webred}{rgb}{0.75,0,0}
\definecolor{webblue}{rgb}{0,0,0.75}
\definecolor{darkblue}{rgb}{0,0,0.7}
\definecolor{dunkelgrau}{rgb}{0.8,0.8,0.8}
\definecolor{lgray}{rgb}{0.95,0.95,0.95}
\definecolor{lgreen}{rgb}{0.95,1.00,0.90}
\definecolor{lblue}{rgb}{0.9,0.95,1.00}
\definecolor{lred}{rgb}{1.00,0.90,0.80}
\definecolor{shadecolor}{rgb}{1.00,0.92,0.82}

\usepackage{iopams}  
\usepackage{graphicx}
\usepackage{cite}
\usepackage[colorlinks=true,linkcolor=darkblue,citecolor=darkblue,urlcolor=darkblue]{hyperref}
\usepackage{hyperref}

\def\endthebibliography{%
  \def\@noitemerr{\@latex@warning{Empty `thebibliography' environment}}%
  \endlist
}

\begin{document}

\title[]{Random Matrix Ensembles in Hyperchaotic Classical Dissipative Dynamical Systems}

\author{J. Odavi\'{c}$^1$ \& P. Mali$^2$}

\address{$^1$Ru\dj er  Bo\v{s}kovi\'{c}  Institute,  Bijeni\v{c}ka  cesta  54,  10000  Zagreb, Croatia\\
$^2$Department of Physics, Faculty of Science, University of Novi Sad, Trg Dositeja Obradovi\' ca 4, 21000 Novi Sad, Serbia}
\ead{jodavic@irb.hr}
\vspace{10pt}
\begin{indented}
\item[]November 2020
\end{indented}

\begin{abstract}
We study the statistical fluctuations of Lyapunov exponents in the  discrete version of the non-integrable perturbed sine-Gordon equation, the dissipative ac+dc driven Frenkel-Kontorova model. Our analysis shows that the fluctuations of the exponent spacings in the strictly overdamped limit, which is nonchaotic, conforms to the \textit{uncorrelated} Poisson distribution. By studying the spatiotemporal dynamics we relate the emergence of the Poissonian statistics to Middleton's no-passing rule. Next, by scanning over the dc driving and particle mass we identify several parameter regions for which this one-dimensional  model exhibits hyperchaotic behavior. Furthermore, in the hyperchaotic regime where roughly fifty percent of exponents are positive, the fluctuations exhibit features of the \textit{correlated} universal statistics of the Gaussian Orthogonal Ensemble (GOE). Due to the dissipative nature of the dynamics, we find that the match, between the Lyapunov spectrum statistics and the universal statistics of GOE, is not complete. Finally, we present evidence supporting the existence of the Tracy-Widom distribution in the fluctuation statistics of the largest Lyapunov exponent. 
\end{abstract}

%\maketitle
\ioptwocol

\section{Introduction }
\label{intro}
For a long time matrices with random entries have been  occupying physicists. The simple reason for this is that many complex and strongly-correlated many-body problems, that can be formulated in terms of ensembles of such random matrices, are found to be analytically tractable due to the underlying symmetries that these matrices entail \cite{Mehta2004}. Starting with the work of Wigner \cite{Wigner_195AD} ensembles of such ``integrable'' Hamiltonians have been shown to naturally emerge in fields such as nuclear physics, disordered systems, string theory, transport phenomena and many others \cite{Weidenmuller_Mitchell_2009,Eynard_Kimura_Ribault_2018,Beenakker_1997}.

Random Matrix Theory (RMT) is a scientific discipline involved in the study of the particular universal features displayed by an ensemble of random matrices when the size of the matrix  $N \!\! \to \! \! \infty$. When the matrix entries are independently and identically distributed (i.i.d.) random variables and the matrices are rotationally invariant three universality classes or ensembles of \textit{correlated} random matrices exist and are designated as Gaussian ensembles. Matrices with $[N \! \times \! N]$ real symmetric random entries are known as Gaussian Orthogonal Ensemble  (GOE), $[N \! \times \! N]$ complex Hermitian  known as Gaussian Unitary Ensemble (GUE), and $[2N  \times  2N]$ self-dual Hermitian matrices are known as Gaussian Symplectic Ensemble (GSE). In the process of diagonalization of the just mentioned matrices the eigenvalues become correlated and each eigenvalue ``feels'' the presence of the neighboring ones leading to the phenomenon of level repulsion \cite{Livan_Novaes_Vivo_2018}. In the asymptotic $N \!\! \to \!\! \infty$ limit the eigenvalue distributions of such ensembles conform to the famous semi-circle law, while the eigenvalue spacings distribution follows the universal Wigner's surmise \cite{Mehta2004,Bronk_1964}.  An interesting variation to the standard Gaussian ensembles include the $[N \times N]$ Wishart matrix $W = X^{\rm T} X$ where $X$ is rectangular $[N' \times N]$ matrix with i.i.d.\ entries, resulting in the Marchenko-Pastur law \cite{Mar_enko_1967}. 
In case the eigenvalues are themselves i.i.d.\ random values (i.e.\ \textit{uncorrelated}) with finite variance, their distribution due to the central limit theorem follows the normal (Gaussian) distribution. Eigenvalue spacings in this case follow the Poisson distribution. To generate such an eigenvalue distribution it is sufficient to diagonalize matrices with i.i.d.\ values sampled from the Gaussian distribution only along the main diagonal, e.g.\ see \cite{Casati_1991}.

Recently, the presence of universal statistics of random matrix ensembles in the Lyapunov exponent spectrum has been demonstrated in the classical limit of the matrix model of D0-branes \cite{Gur-Ari_Hanada_Shenker_2016,Hanada_Shimada_Tezuka_2018} with conservative dynamics. The Lyapunov exponents measure the average exponential rate of the divergence of neighboring orbits in phase space. They are considered an indispensable tool for detecting the presence of chaos in dynamical systems \cite{Skokos_Gottwald_Laskar_2016}. The Lyapunov exponent distribution in the above mentioned  supersymmetric model, used by the string theory community, was shown to follow the semi-circle law. Furthermore, in the Kuramoto model of $N$ oscillators with variable coupling matrix similar findings have been reported, where the presence of Poisson or Wigner surmise distribution in the Lyapunov exponent spacings is taken as an indicator for  synchronization behavior \cite{Patra_Ghosh_2016}. Historically, authors in \cite{Ahlers_Zillmer_Pikovsky_2001} were the first to show the existence of neighboring Lyapunov exponents repulsion, hinting at the similarities in the behavior of Lyapunov exponents and the Gaussian RMT ensembles. 

Motivated by the above-mentioned findings we proceed to show the existence of correlated random matrix ensemble features in a typical condensed-matter model, the classical nonlinear and hyperchaotic dynamical system with \textit{dissipative} dynamics, the ac+dc driven Frenkel-Kontorova model. Moreover, within the same model and for a particular choice of parameters when the system is nonchaotic we show the presence of the Poisson statistics in the Lyapunov exponent spacings. We choose to work with this model because of its clear physical interpretation and enormous applicability in various existing physical systems \cite{OBBook}. This model represents an appropriate theoretical framework for the description of charge- and spin-density wave transport, irradiated Josephson-junction arrays, and driven colloids \cite{Thorne_Hubacek_Lyons_Lyding_Tucker_1988,ACDS,Juniper_Straube_Besseling_Aarts_Dullens_2015},  Later, in our paper, we show that the considered model in the regime where its Lyapunov exponents exhibit repulsion is equivalent to a realistic one-dimensional parallel array of Josephson junctions having the stripline geometry \cite{Pfeiffer_Abdumalikov_Schuster_Ustinov_2008}.  

The Lyapunov exponents are typically defined as the singular value decomposition values of the Jacobians of the linearized dynamics. The Jacobian that governs the behavior of the perturbed trajectories at the end of the time evolution of a dynamical system is composed out of the product of such Jacobians at previous time steps \cite{Wolf_Swift_Swinney_Vastano_1985,PikovskyPoliti2016}. From this point of view the eigenvalue statistics of the product of random matrices and Lyapunov exponent statistics are closely related; see Ref.\ \cite{CrisantiPaladinVulpiani1993} and references therein. Making use of this connection the computation of the complete Lyapunov exponent spectrum has been approximately performed for the Fermi-Pasta-Ulam chain of oscillators \cite{EckmannWayne1988}. Our work, on the other hand, is focused on identifying different dynamical regimes in the spatially extended dissipative and ac+dc driven Frenkel-Kontorova model from the Lyapunov spectrum. In this effort, we compare the Lyapunov exponents' fluctuations to the predictions of RMT.  

The paper is organized in the following way. In Sect.\ \ref{model} we present the model and discuss the interesting regimes and model-specific nomenclature used in the paper. Next, in Sect.\ \ref{spectrum} we discuss how the Lyapunov exponent spectrum is defined and its particular importance. In Sect.\ \ref{results} we present the results of our study using techniques from both RMT and nonlinear dynamics community. We conclude in Sect.\ \ref{conclusion}.

%================================================================================================================================
\section{Model and method}
\subsection{The model}\label{model}

Standard Frenkel-Kontorova (FK) model represents a chain of harmonically interacting identical particles with positions $\{u_i\}$
 subjected to the sinusoidal substrate potential with amplitude $K$
\begin{equation}
V(u)=\frac{K}{4\pi^2}\big[1-\cos(2\pi u)\big],
\end{equation}
and it is defined by the Hamiltonian
\begin{equation}
H=\sum_i \Big(\frac{m}{2}\dot{u}_i^2+\frac{1}{2}(u_{i+1}-u_i - a_{0})^2+V(u_i) \Big), \label{model1}
\end{equation}
where $m$ represents the mass of the point-like particles each indexed by subscript $i$, while $a_{0}$ is the equilibrium distance of the inter-particle potential. After proper renormalization of model parameters and by neglecting the discreteness effects, the standard FK model reduces to the well-studied integrable sine-Gordon (sG) equation  \cite{OBBook}.  sG is applicable in a wide range of physical systems from Josephson junctions (JJs) to gravity and high-energy physics \cite{OBBook,Cuevas-Maraver_Kevrekidis_Williams_2014}. Systems that are described by the sG equation are known to host excitations such as topological solitons (kinks) and dynamical solitons (breathers) and attract a lot of attention from physicists despite/for being integrable. The sG equation can be extended in different ways. For instance, the common and useful extension is the perturbed sG equation. Depending on the physical system and interpretation the perturbation is typically a damping term that breaks the integrability. A particular version of the perturbed sG equation that includes both damping and a driving term has been successful in describing the long JJs \cite{McLaughlin_Scott_1978}.
\begin{figure}[] %\bigskip
\includegraphics[width=7cm]{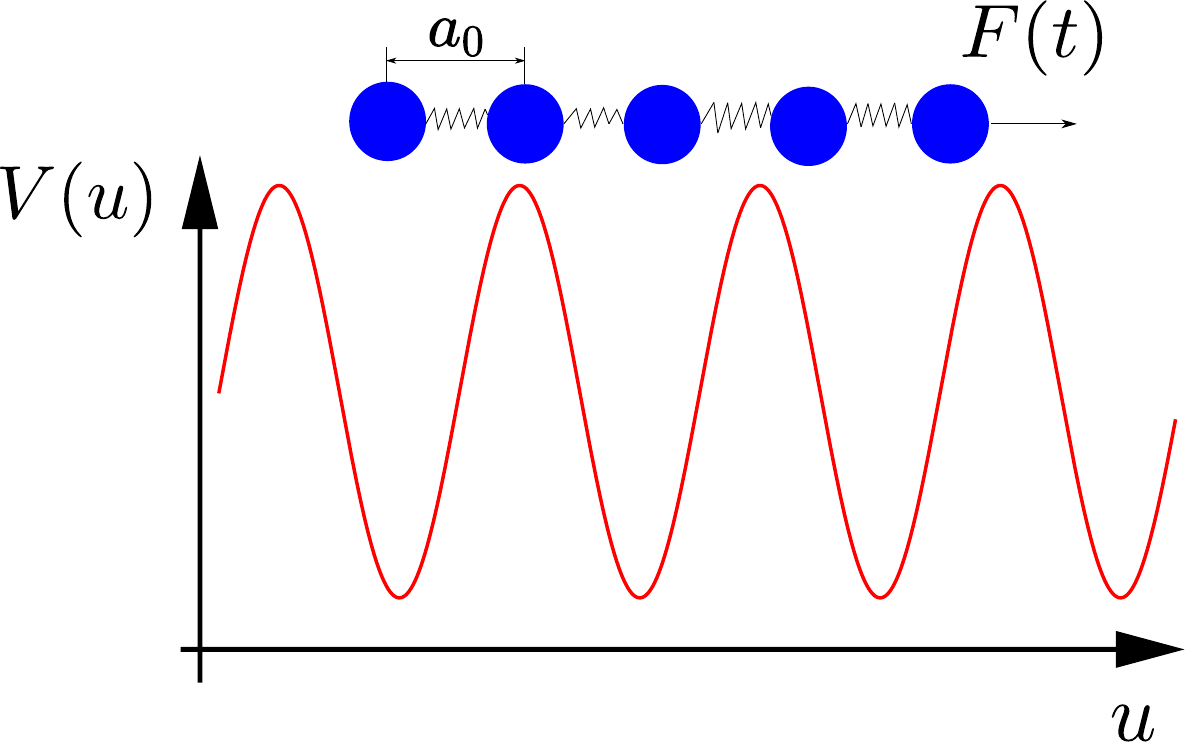}
\centering
\caption{\label{Fig1}(Color online) Sketch of the ac+dc driven Frenkel-Kontorova model defined with Eq.\ (\ref{model1}) and equations of motion given in Eq.\ (\ref{model2}). The harmonically interacting particles are featureless point-like masses.}
\end{figure}

In our work, we focus on the dissipatively driven FK model, i.e.\ discretized version of the perturbed sG equation, which is also non-integrable and exhibits more complex dynamics compared to sG equation and standard FK model. Such a model is used to describe phenomena in many different physical systems \cite{OBBook,ACFK,FlorAP,Tosati,Ustinov}. Equations of motion of the dissipatively driven FK model of $N$ identical particles have the following form
\begin{equation}
m\ddot{u}_i=u_{i+1}+u_{i-1}-2u_i-\frac{\partial V}{\partial u_i}-\dot{u}_i+F(t), \label{model2}
\end{equation} 
where $\dot{u}_i$ defines the dissipative term, $i$ labels particles $i=1,...,N$, and $F(t)$ is the driving force  chosen in the form $F(t)=F_{\rm{dc}}+F_{\rm{ac}}\cos(2\pi\nu_0 t)$ (see Fig.\ \ref{Fig1}). We impose cyclic boundary conditions by $u_0=  {\rm mod}(u_N,N\omega)$, where $\omega$ represents the interparticle average distance, i.e.\ average number of particles per substrate potential well \cite{FlorAP,Odavic}. In literature, different regimes have been identified to exist in this model such as $m \to 0$ limit of Eq.\ (\ref{model2}) which defines the strictly overdamped limit \cite{Falo}, the overdamped regime for which $0 < m \leq \frac{1}{4(2+K)}$, and the underdamped regime otherwise \cite{Baesens_MacKay_2003}. To integrate the system of Eqs.\ (\ref{model2}) we use standard techniques from \cite{solver} with the time step $\nu_{0}^{-1}$ and relative tolerance of $10^{-6}$.

Due to the exertion of the force term $F(t)$, the dissipative FK model undergoes a dynamical phase transition from a pinned to the sliding particle motion regime \cite{Reichhardt_Reichhardt_2016,Odavic}. The study of the critical force dynamics is an active field of research related to the description of different interesting phenomena in tribology, traffic flow of cars on the road, and many others \cite{OBBook}. In our work, we primarily investigate the physics of the sliding regime where the system's response is strongly nonlinear and the dynamics is richer.

In the underdamped regime, the dissipatively driven FK model exhibits sensitivity to initial conditions \cite{ACm}, i.e.\ chaotic dynamics, whereas in the strictly overdamped limit the system is nonchaotic \cite{ACDS,Odavic}. In the following section, we describe how the presence of chaos is quantified using Lyapunov exponents.

\subsection{Spectrum of Lyapunov exponents}\label{spectrum}
\begin{figure}[] %\bigskip
\includegraphics[width=\columnwidth]{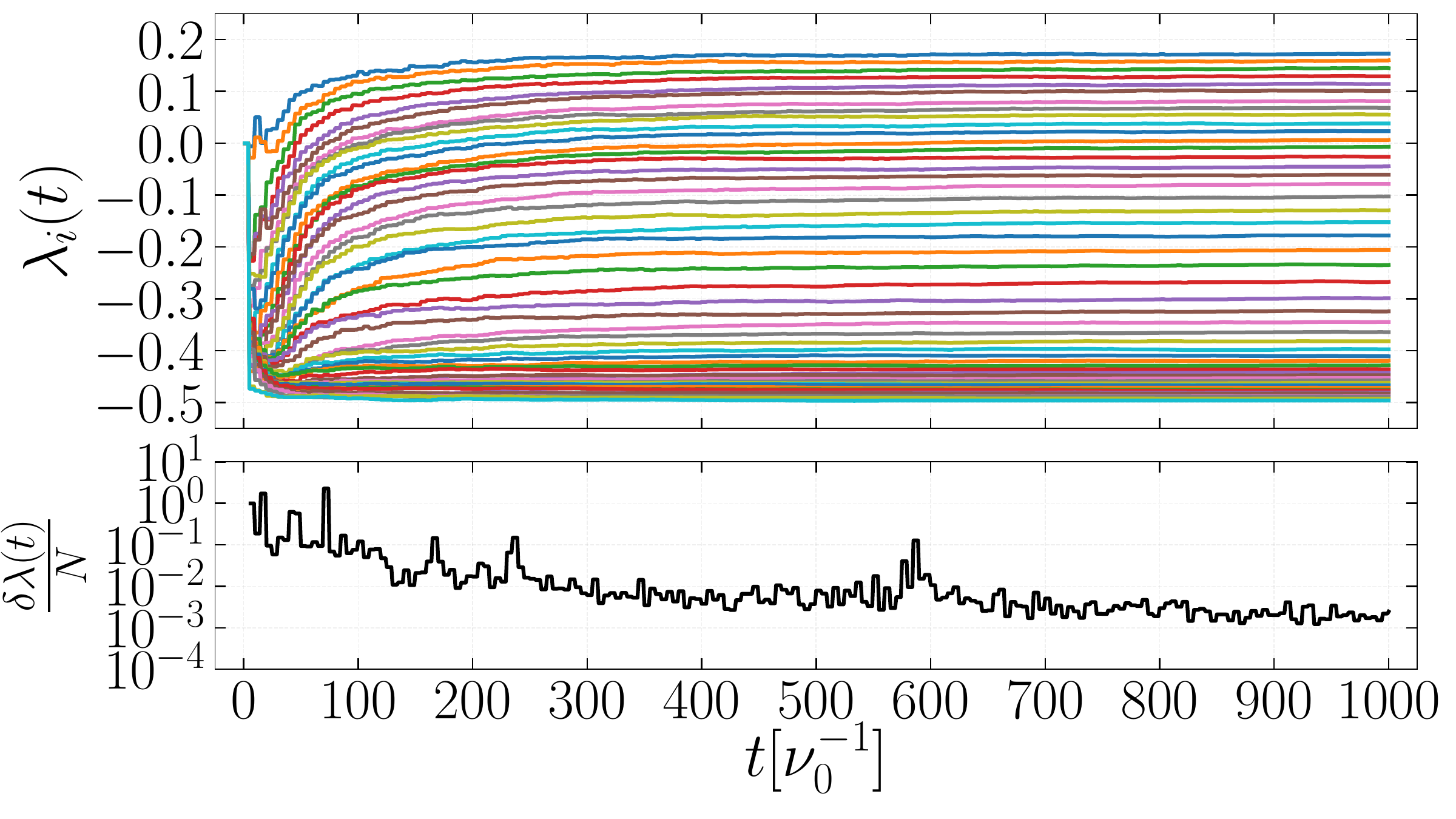}
\centering
\caption{\label{Fig2}(Color online) Time evolution of Lyapunov exponents $\lambda_i(t)$ (top panel) and from Eq.\ (\ref{error}) the average relative error $\frac{\delta \lambda}{N}$  (bottom panel) for dissipative driven FK model with parameters $N=50$, $\omega=1/2$, $F_{\rm{ac}}=0.2$, $\nu_0=0.2$, $K=4$, $m=1$, $F_{\rm{dc}}=0.17$. }
\end{figure}
Lyapunov exponents (LEs) are an essential diagnostic tool for the stability of attractors and the presence of deterministic chaos in dynamical systems. They quantify the average exponential rates of divergence (or convergence) of neighboring orbits in phase space \cite{Skokos_Gottwald_Laskar_2016}. An ordered set of LEs $\{ \lambda_{1}, \lambda_{2}, ..., \lambda_{n} \}$ forms the spectrum, where the cardinality of the set $n$ matches the number of system's degrees of freedom. If the system has at least one positive LE the system is chaotic, and if we order the spectrum $\lambda_{1} \geq \lambda_{2} \geq ... \geq \lambda_{n}$ the maximal (largest) Lyapunov exponent is then $\lambda_{\rm max} = \lambda_{1}$. Moreover, if multiple exponents are positive, the dynamics are designated as \textit{hyperchaotic} \cite{Rossler_1979,Matsumoto_Chua_Kobayashi_1986}.  

 To numerically estimate the exponents  a set of linearized equations  with perturbed initial conditions is solved. If $u_{i} (t)$ are the particle trajectories, the perturbed trajectories $\delta u_{i} (t)$ are determined by the successive application, at each time step, of the linear propagators $J_{i,j}$ (the Jacobian) to the initial perturbed positions $\delta u_{i} (0)$ as  
\begin{equation}
\delta u_{i} (t) = \sum\limits_{j} J_{i,j} (t,0) \delta u_{i} (0). \label{jacobian}
\end{equation}
The finite-time LEs are defined as
\begin{equation}
\lambda_{i} (t) = \lim_{ \parallel \delta \parallel \to 0} \frac{1}{t} \ln \frac{\parallel \! \delta u_{i} (t) \! \parallel}{\parallel \! \delta u_{i} (0) \!\parallel},
\end{equation}
and also represent the eigenvalues of the Jacobians. In our work, we examine the fluctuations of finite-time LEs around their converged (or ``asymptotic'') limit and after the system has reached the steady-state dynamical regime.

To compute the LE spectrum we employ the algorithm from Ref. \cite{Wolf_Swift_Swinney_Vastano_1985}, first proposed in Refs.  \cite{Benettin_Galgani_Giorgilli_Strelcyn_1980,Shimada_Nagashima_1979}. 
This particular algorithm was shown to be sufficiently accurate in the computation of the LEs and the associated eigenvectors for a multitude of systems, both conservative and dissipative \cite{Ramasubramanian_Sriram_2000}.

As a criterion when the LE spectrum is numerically converged we take the following average relative error estimate
\begin{equation}
\frac{\delta \lambda (t_{j})}{N} =\frac{1}{N}\sum^N_{i=1} \Bigg\vert \frac{\lambda_i(t_j)-\lambda_i(t_{j-1}) }{\lambda_i(t_j)} \Bigg\vert
< 5 \cdot 10^{-3}, \label{error}
\end{equation}
where $t_{j} = 5 j \nu_{0}^{-1} $. This particular convergence criterion choice is made empirically by minimizing the trade-off between execution costs and precision. However, enforcing higher precision does not change our main results  but rather changes the scale of the phenomenology we report on. In Fig.\ \ref{Fig2} we illustrate how the LE spectrum evolves and converges over time for a typical parameter set. In the top panel, we observe the individual exponents reach their ``asymptotic'' values in a consistent manner which is captured by the relative error estimate $\delta \lambda /N$ (bottom panel). For the model under study, we performed the Gram-Schmidt orthonormalization of the linearized basis at every time step \cite{Wolf_Swift_Swinney_Vastano_1985}. The difference with the usual application of local (finite-time) LE and the way we calculated the exponents (see Eq. 6 for comparison in Ref.\ \cite{Prasad}) is that our evaluation interval is very short, i.e.\ of the order of the time step we take. As we take the LE measurements in the steady-state dynamical regime the evaluation interval can indeed be this short. 

\begin{figure*}[]
  \includegraphics[width=8.9cm]{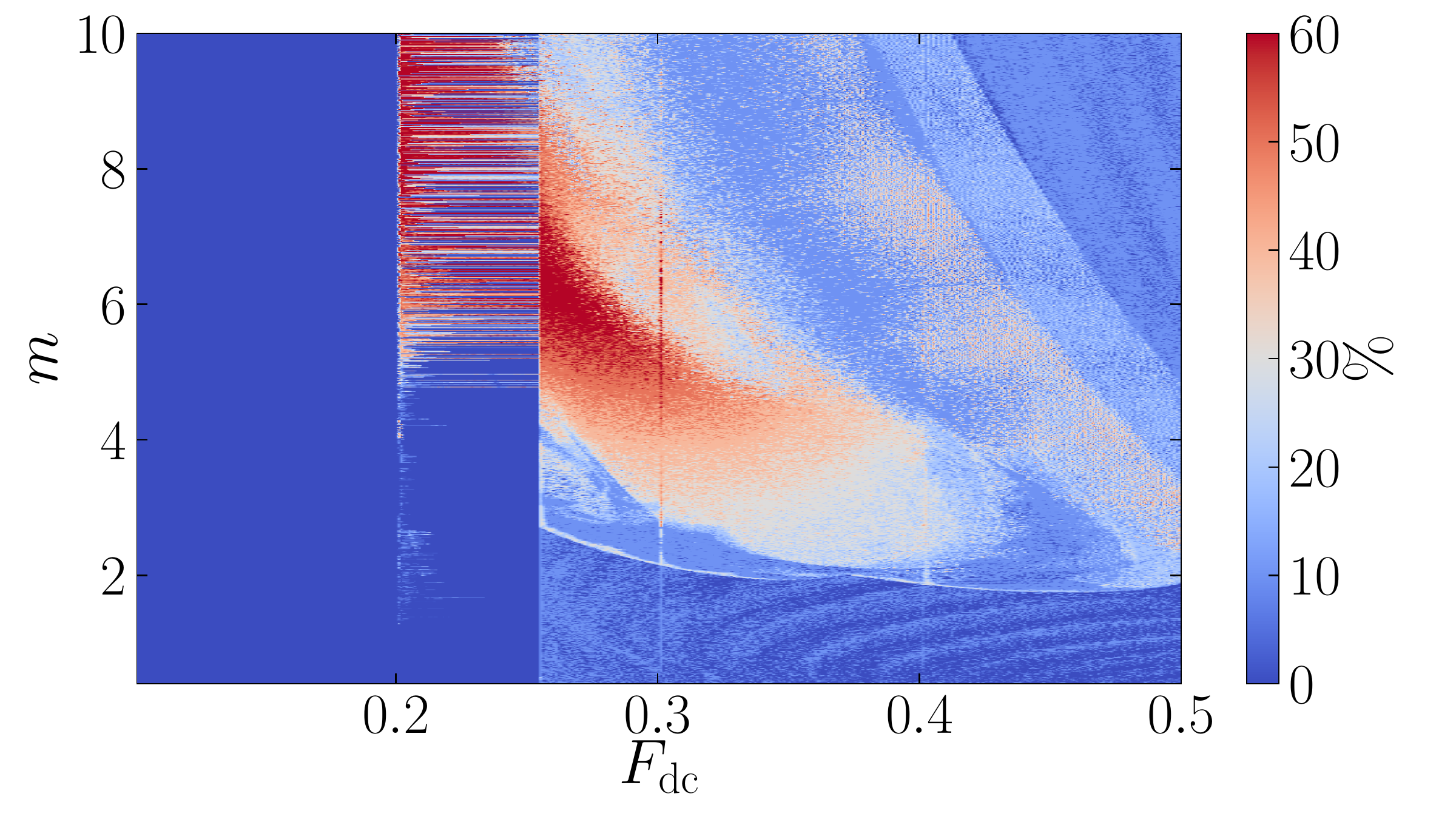}\hfill
  \includegraphics[width=8.9cm]{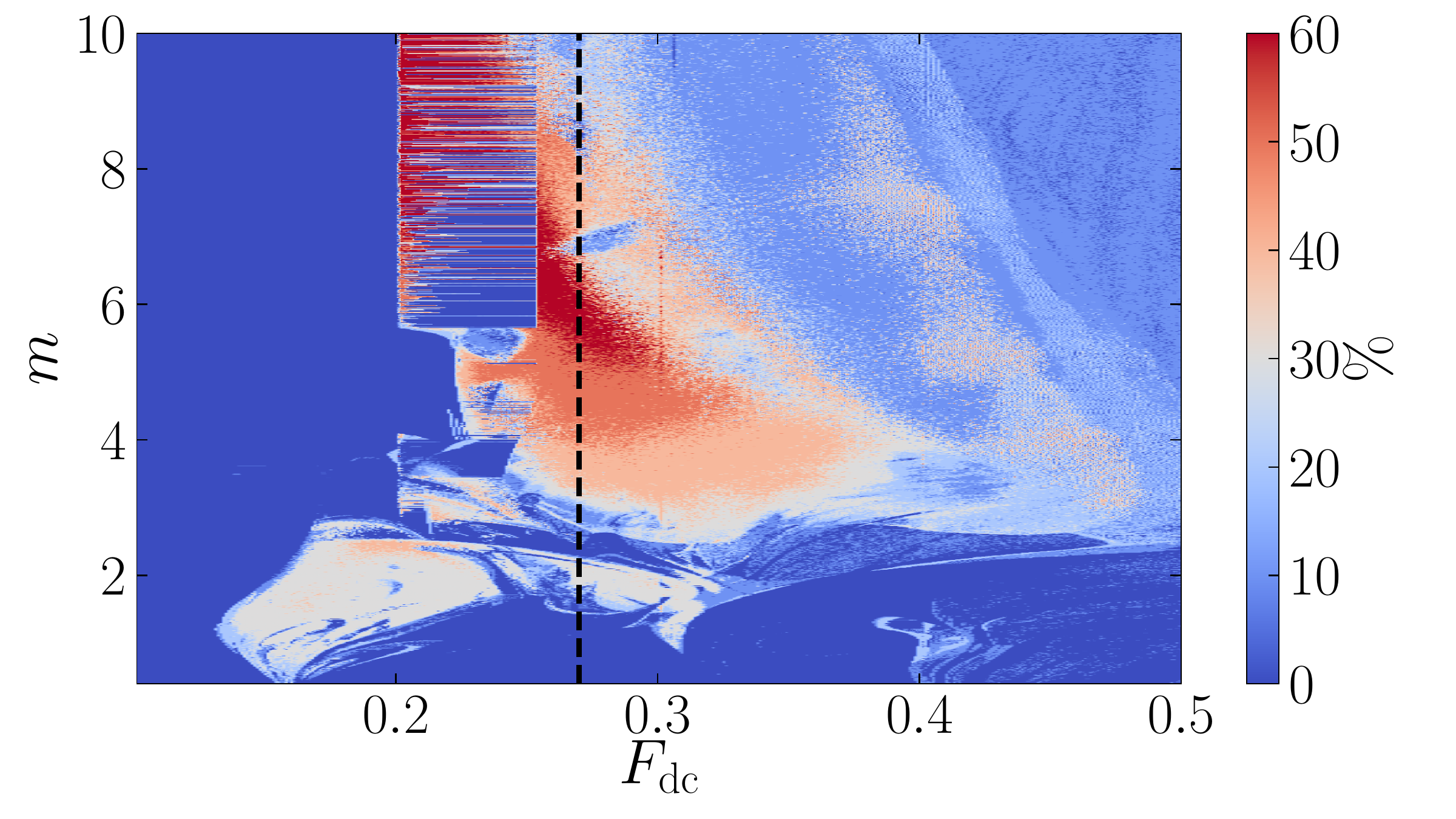}
  \caption{\label{Fig3}(Color online) Percentage of positive LEs as a function of the dc force $F_{\rm{dc}}$ and the mass of particles $m$, for parameters $N=10$, $\omega=1/2$, $\nu_0=0.2$, $K=4$. Force and mass step used $\Delta F = 10^{-3} $, $\Delta m = 5 \cdot 10^{-3}$. Left panel: dc system with $F_{\rm {ac}} = 0$; Right panel: ac+dc system with $F_{\rm {ac}} = 0.2$ where the vertical black dashed line is at $F_{\rm dc} = 0.27$. Parameter regimes along the dashed are examined later(see Sect.\ \ref{resultsA}).  }
\end{figure*}

In the underdamped regime the equations of motion are of second-order (see Eq.\ (\ref{model2})), and to integrate them we instead solve two sets of $N$ first-order differential equations. Therefore, the size of the Jacobian matrix from Eq.\ (\ref{jacobian}) is $ \big[ 2N + 1, 2N + 1 \big]$, where the last equation is reserved for time. This amounts to simultaneous solving $(2 N + 1) \times (2 N + 1)$ linear and $2 N + 1$ nonlinear first-order differential equations.  In our study of the underdamped regime, we exclusively focus on $N$ largest exponents out of the total $2N$ number of exponents. In Appendix \ref{appendixA} we motivate this choice. However, in the strictly overdamped case, the total number of first-order equations (linear + nonlinear) instead is $ (N + 2) \times (N + 1)$ and requires a separate implementation. Moreover, in this case, the particles are inertialess and half of the available phase space directions are neglected making the total number of exponents $N$. We verified that by working with Eq.\ (\ref{model2}) and slowly approaching $m \to 0$ we can obtain the results of the strictly overdamped limit with a precision allowed by our integration routines (see also Sect.\ \ref{results} and Fig.\ \ref{Fig4}).

The magnitude of LEs measures the rate at which the system becomes unpredictable and how fast the information about the initial state gets scrambled \cite{Gur-Ari_Hanada_Shenker_2016}. The largest LE can be related to dynamics reversibility and Loschmidth echo, e.g.\ \cite{JalabertPastawski2001,Tarkhov_Wimberger_Fine_2017,Veble_Prosen_2004}. In the strictly overdamped limit of the FK model, the largest LE at vanishing dc driving was shown to be in close relation to the critical depinning force \cite{Odavic}. Moreover, in the JJ arrays systems, the critical depinning force is analogous to the critical current \cite{Reichhardt_Reichhardt_2016}, therefore the study of the largest LE is of principal importance. While in the Kuramoto model the characteristics of the LE spectrum are used to distinguish between different dynamical regimes \cite{Patra_Ghosh_2016}. Similarly, we identify GOE features (see Sects.\ \ref{resultsA} \ref{resultsB}, \ref{resultsD} and \ref{resultsF}) in the dissipatively driven FK model and show how from the LE spectrum we identify several different dynamical regimes (see Sect. \ref{resultsC}).

%==================================================================
\section{Results}
\label{results}

The fact that the dissipatively driven FK model exhibits hyperchaotic features, i.e.\ a dynamical regime with more than one positive LE, was already highlighted in \cite{ACm}. In this paper, we perform a more in-depth study of this interesting phenomenon and present new results related to the complex dynamical landscapes emerging in this at first glance simple model. 

In Fig.\ \ref{Fig3} we show the heatmap of the results of a scan over the mass $m$ and dc force parameter $F_{\rm dc}$. By measuring the percentage of positive LEs in the spectrum we can identify several different parameter regimes that lead to dynamics with strong chaoticity. In the left panel, the results for a dc system ($F_{\rm ac} = 0$)  and in the right one ac+dc system ($F_{\rm ac} = 0.2$)  are presented. In both cases, the dynamical landscapes exhibit large periodic phases (dark blue regions) with smaller chaotic windows (light blue to red regions).

\begin{figure}[] %\bigskip
\includegraphics[width=\columnwidth]{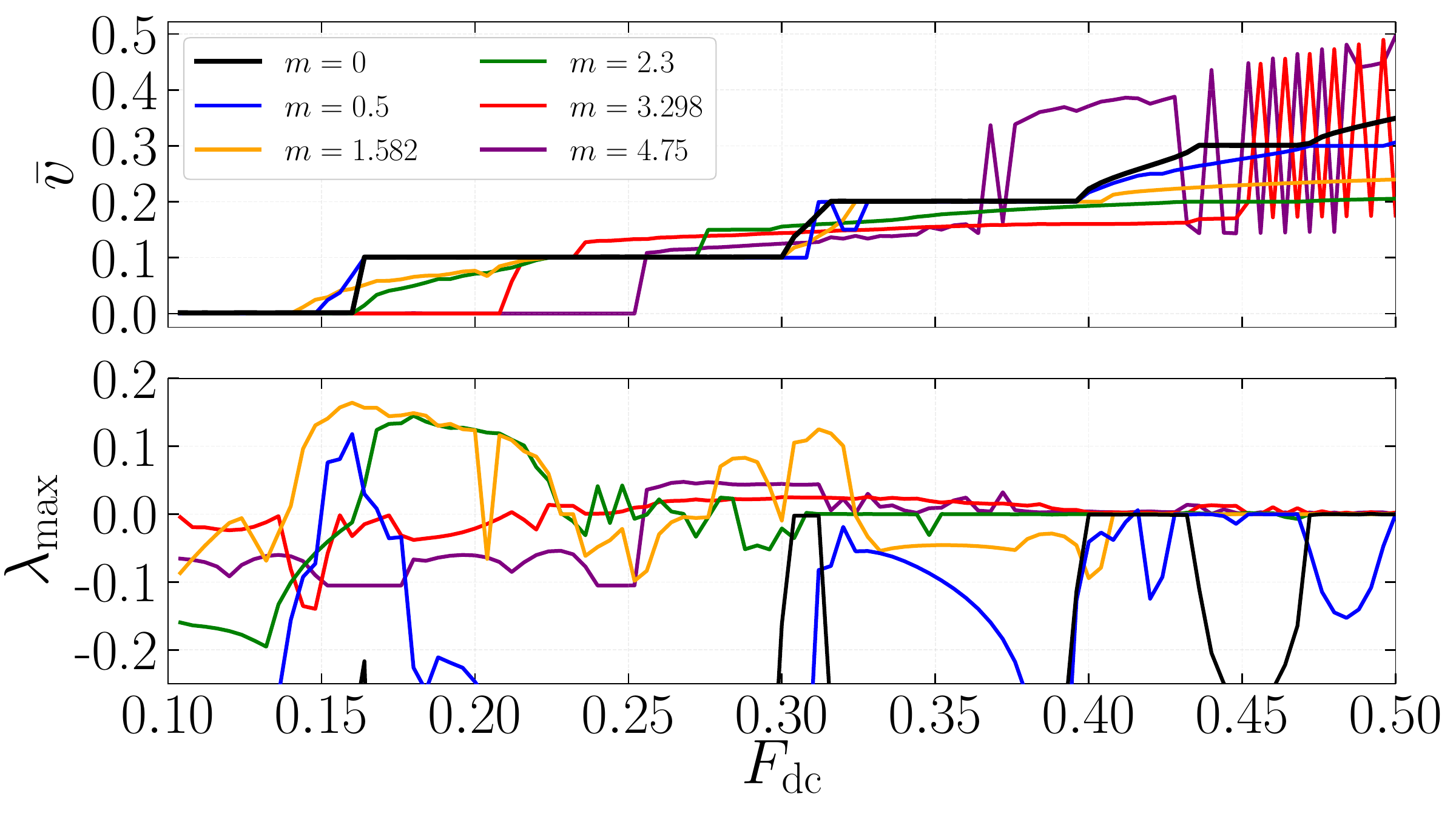}
\centering
\caption{\label{Fig4}(Color online) Average velocity $\bar{v}$ (top panel) and maximal (largest) Lyapunov exponent $\lambda_{\mathsf{max}}$ (bottom panel) as a function of dc driving force $F_{\rm{dc}}$ for five different masses  and strictly overdamped limit. Parameters used: $N = 10$, $\omega = 1/2$, $F_{\rm ac} = 0.2$, $\nu_{0} = 0.2$, and $K = 4$ with dc force step $\Delta F_{\rm dc} = 4 \cdot 10^{-3}$. }
\end{figure}

We now additionally comment on the consistency of our calculations and numerical implementations. In particular, from \cite{ACm} we know that whenever $\lambda_{\rm max} > 0$ the system typically exhibits collective motion measured by the response function $\bar{v}$ which is defined as 
\begin{equation}
\bar{v} = \Big\langle \!\! \Big\langle \dot{u}_{i}(t) \Big\rangle \!\! \Big\rangle_{T,N} = \lim\limits_{T \to \infty} \frac{1}{T N} \sum\limits_{i = 1}^{N} \int\limits_{t_{\rm s}}^{t_{\rm s} + T}  \dot{u}_i (t) {\rm d} t  \neq 0,
\end{equation}
where $t_{\rm s}$ is the elapsed transient time, i.e.\ time needed for the system to reach the steady-state. Knowing this, we can be confident that our numerical integration of  Eq.\ (\ref{model2}) and LE spectrum computation from Sect.\ \ref{spectrum} are correctly implemented. More specifically, we know that in the ac+dc strictly overdamped limit the dynamical phase transition from the pinned to the sliding particle motion regime for the parameters specified in the captions of Fig.\ \ref{Fig3} happens at around $F_{\rm dc} \!\! \sim \! 0.16$ \cite{Odavic,Falo} (see also the top panel of Fig.\ \ref{Fig4}). The results for the underdamped system in the right panel of Fig.\ \ref{Fig3} show that such transition indeed happens in the $m\! \to \!0$ limit where the chaotic patch shrinks towards this particular critical depinning force. 

In both the dc and ac+dc case a regime with strong intermittent chaotic behavior is present. Alternations between chaotic and periodic windows happen for $0.2 < F_{\rm dc} < 0.26$ with very small change in the mass parameter. This means that pinning-to-sliding transition in this model is more complex than previously thought. Therefore, going further with our analysis we shall focus on parameters that lead to dynamics deep inside the sliding regime where such intermittent behavior does not dominate the physics and stable phases are present, i.e.\  same colored islands in the heatmap of Fig.\ \ref{Fig3}. The ac+dc case exhibits larger and more robust islands. Due to this fact the model that includes both  the ac and dc driving is the subject of our further investigation.

To illustrate the different dynamical regimes present in the driven FK model, in Fig.\ \ref{Fig4}, we plot the response function $\bar{v}(F_{\rm dc})$ (upper panel) and the largest Lyapunov exponent $\lambda_{\rm max}(F_{\rm dc})$ (lower panel). The response function and also the LEs, with the decrease of the mass parameter $m \! \to \! 0$, converges towards the results for the strictly overdamped model and are consistent with standard literature reference \cite{Falo}. We note that we neglect any hysteresis effects in the response function by independently running the calculations for each particular $F_{\rm dc}$ \cite{ACm}. 

We checked that for larger driving frequency, e.g. $\nu_{0}  = 0.4$, the chaotic behavior is more pronounced, and chaotic regions are larger. According to \cite{Kautz_Monaco_1985} and the author's work on the related model of JJ array, there exists only a particular frequency parameter window where chaotic behavior is present and the size of this window is related to the McCumber's parameter. Due to the obvious equivalence of the models, in our model,  the size of this window is related to the mass $m$. Moreover, performing calculations for larger ac force leads to hyperchaotic islands that are already present for smaller $F_{\rm ac}$. Additionally, the coupling between junctions in the related model of JJ arrays was shown to play an important role in inducing chaotic behavior \cite{Shukrinov_Azemtsa-Donfack_Botha_2015}.

To answer commonly asked questions regarding the commensurate dynamics of the FK model, a small particle number is typically  used \cite{OBBook,ACFK}. More specifically, for fixed $\omega$ the $N \leq 10$ is shown to be sufficient to measure the systems response function $\bar{v} (F_{\rm dc})$ and if the system is chaotic or not \cite{ACDS} (also see Fig.\ \ref{Fig4}), and these results would carry over and be valid for $N  >  10$ cases. However, deep inside the underdamped regime, the physics (for the considered system sizes) changes with the particle number. In particular, the percentage of positive LEs changes. Luckily, the change in the number of positive LE exponents with $N$ is gradual. That is why we compute the heatmap with only $N = 10$, presented in Fig.\ \ref{Fig3}, with the expectation that similar (within a few percent) chaotic content will be present in larger systems.

The presence of hyperchaoticity implies that information about the initial state of the system gets scrambled along with several directions in the phase space and the existence of a higher-dimensional attractor structure. The LE spectrum and the associated attractor is characterized by its fractal dimension, correlation exponent, and information dimension. Quantities such as the Kaplan-York dimension  can be computed and measure the information content of a dynamical model, whereas the sum of positive LEs is known to be related to Kolmogorov-Sinai entropy and the entropy growth rate \cite{Wolf_Swift_Swinney_Vastano_1985}. In this paper, we do not try to answer questions about information and entropy but rather focus on the statistics of the LE spectrum, which we show to exhibit certain features of the Gaussian random matrix ensembles \cite{Mehta2004,Wigner_195AD,Weidenmuller_Mitchell_2009,Eynard_Kimura_Ribault_2018,Beenakker_1997,Livan_Novaes_Vivo_2018}.

\begin{figure}[] %\bigskip
\includegraphics[width=\columnwidth]{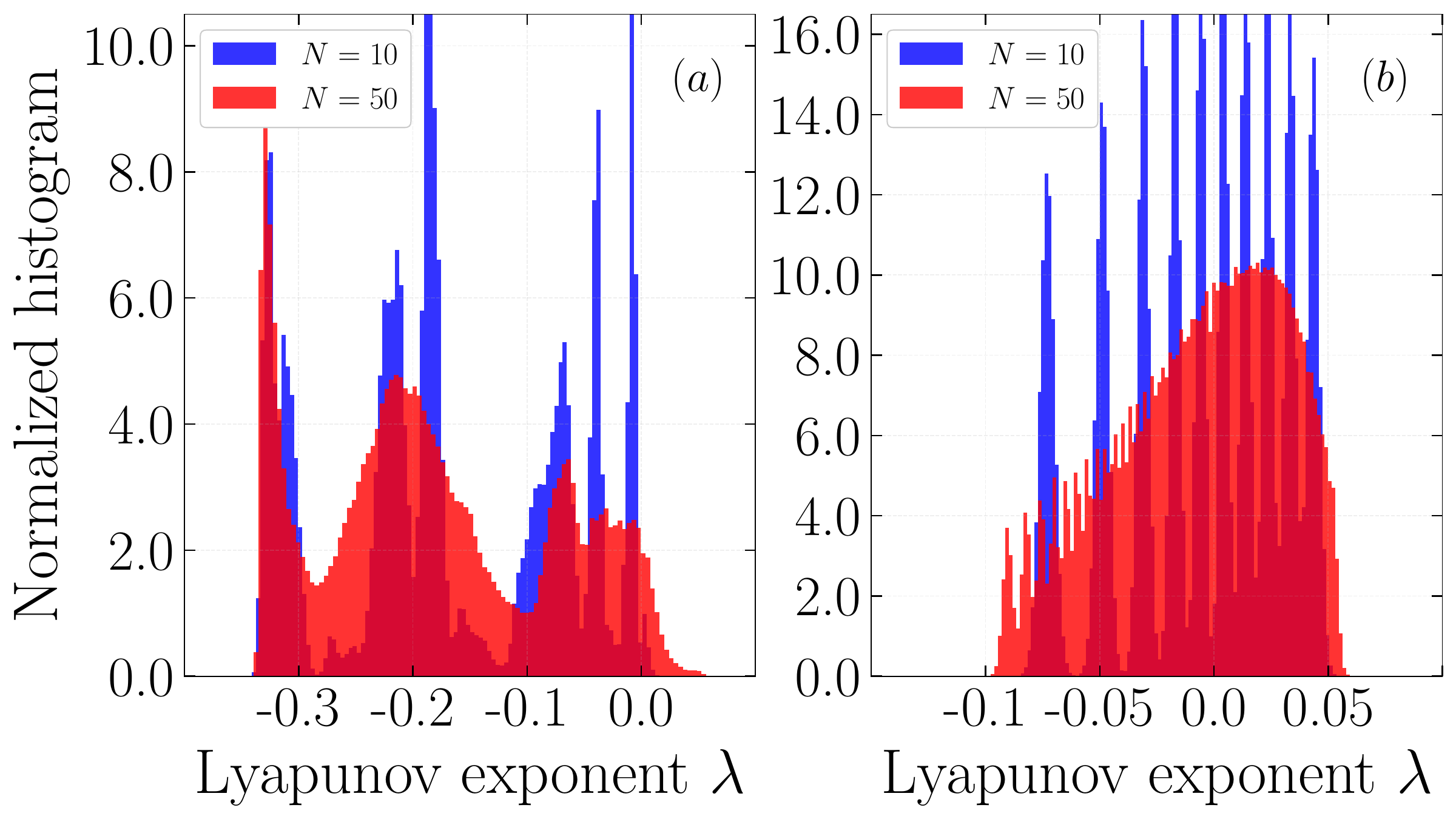}
\centering
\caption{\label{Fig5}(Color online) Histogram of LE spectrum for different particle number $N \! = \! 10, 50$  for large number of realization $M \! = \! 10^5, 10^{4}$, respectively. Number of histogram bins is 100. In a.) $m \! = \! 1.582$ and b.) $m \! = \! 4.653$ while keeping $F_{\rm dc} \! = \! 0.27$. Remaining parameters used stated in Fig.\ \ref{Fig4}. Further details are given in Sect.\ \ref{resultsA}.    }
\end{figure}

To that end, next, we comment on the consistency of the large degree of freedom limit $N  \to  \infty$ of this discrete model. Taking this limit is equivalent to the investigation of the perturbed sG equation itself. To achieve a reasonably ``smooth'' LE spectrum distribution a sufficiently large number of particles have to be employed. This is obvious from the plot of the LE spectrum histogram in Fig.\ \ref{Fig5} for increasing particle number $N$. For $N = 10$ the histograms and the corresponding distributions are not smooth enough to be compared to the explicit asymptotic results in RMT.

We found that already for $N \! = \! 50$ the LE spectrum contains all the main features which are present for much larger particle numbers with similar statistics. For example, in Fig.\ \ref{Fig5}, only small fluctuations around the left tail of the distribution can be observed when the particle number is doubled. These fluctuations influence the output of the unfolding procedure discussed in Sect.\ \ref{resultsB}, but the cost of integrating a system with $N$ degrees of freedom with $2N$ dynamical variables scales as $\sim \!\! N^{2}$ and the added precision does not warrant the steep computational cost. Therefore, for our scientific purposes of pointing out the similarities between the LE spectrum of a dissipative system and RMT ensembles a lower precision can be justified. Nevertheless, to decrease the finite-size effects even further in the following we typical use $N \! = \! 100$ particles.
   
\subsection{Normalized Lyapunov spectrum density} \label{resultsA}

\begin{figure}[] %\bigskip
\includegraphics[width=\columnwidth]{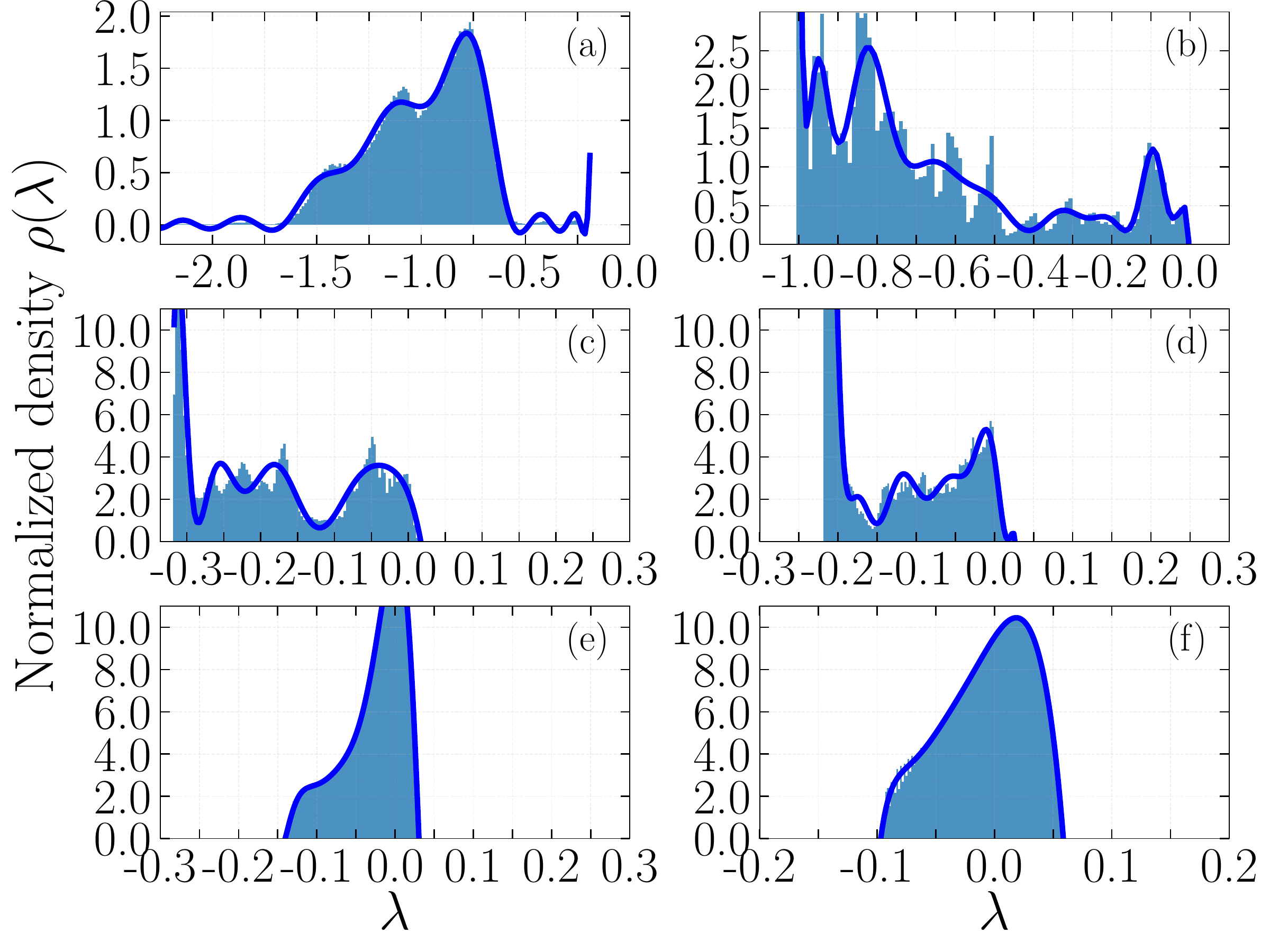}
\centering
\caption{(Color online) Normalized LE spectrum density at $F_{\rm dc} = 0.27$, $\omega = 1/2$, $F_{\rm ac} = 0.2$ and $\nu_{0} = 0.2$ for different masses: $({\rm a})$ $m = 0$, $N = 400$ and $M = 790 $, ${\rm (b)}$ $m = 0.5$, $N = 100$ and $M = 2265$, ${\rm (c)}$ $m = 1.582$, $N = 100$ and $M = 4607$, ${\rm (d)}$ $m = 2.3$, $N = 100$ and $M = 610$, ${\rm (e)}$ $m = 3.298$, $N = 100$ and $M = 4189$ and finally ${\rm (f)}$ $m = 4.75$, $N = 100$ and $M = 66370$. The corresponding averages are: $({\rm a})$ $\bar{\lambda} \simeq  -0.995$, ${\rm (b)}$ $\bar{\lambda} \simeq -0.684$, ${\rm (c)}$ $\bar{\lambda} \simeq -0.177$, ${\rm (d)}$ $\bar{\lambda} \simeq -0.133$, ${\rm (e)}$ $\bar{\lambda} \simeq -0.026$ and ${\rm (f)}$ $\bar{\lambda} \simeq -0.006$.  The dark blue lines are the fits obtained via Eq.\ (\ref{fit}) and represent the PDFs of the respected spectrum. Number of histogram bins is 100. \label{Fig6}  }
\end{figure}

In Fig.\ \ref{Fig6} we present the results for the normalized LE spectrum as a function of the mass term $m$. We fixed the driving force at $F_{\rm dc} \! = \! 0.27$  and $N \! = \! 100$ with the remaining parameters given in the captions of Fig.\ \ref{Fig3} (except for the strictly overdamped limit for which $N \! = \! 400$). For this parameter choice, the system exhibits collective motion $\bar{v}\neq 0$ as observed from the upper panel of Fig.\ \ref{Fig4}. 

  The histograms (blue rectangles) in Fig.\ \ref{Fig6} are generated for a large number of realization $M$ at $t=0$, and after the system has reached the steady-state and fulfilled the convergence criterion from Eq.\ (\ref{error}). The initial conditions are chosen such that the particle positions are Gaussian distributed with mean $\mu = N/2$ and standard deviation $\sigma$ is selected such that the particles symmetrically fall into the domain $[0,N \omega]$. Additionally, the initial velocities of the particles are set to be equal to zero. Drawing the initial particle positions from the uniform distribution on the same domain yields equivalent results for the LE spectrum.

The dark blue lines in the figures represent the fit of the histogram data to a higher-order polynomial 
\begin{equation}
\rho (\lambda, \bar{\lambda}) = \sum_{k = 0}^{k_{\rm max}}  a_{k} (\lambda - \bar{\lambda})^{k}, \label{fit}
\end{equation}   
where $k_{\rm max}$ is chosen depending on the details of the spectrum. In particular, the more local maxima the distribution has, the larger $k_{\rm max}$ is required. For example, for $m = 4.75$ depicted in Fig.\ \ref{Fig6} $({\rm f})$, the choice $k_{\rm max} = 10$ accurately approximates the distribution \cite{Hanada_Shimada_Tezuka_2018}. The fit is centered around $\bar{\lambda}$ which is defined as the ensemble average
\vspace{-9pt}
\begin{equation}
\bar{\lambda} = \Big\langle \!\! \Big\langle \lambda_{i,j} \Big\rangle \!\! \Big\rangle_{N,M}  = \frac{1}{N M} \sum\limits_{j = 1}^{M} \sum\limits_{i = 1}^{N}  \lambda_{i,j},
\end{equation}
where here $j$ specifies the initialization. The fitted lines do not capture the extremely detailed structure of the spectrum for smaller masses, e.g.\ $({\rm b})$ and $({\rm c})$ in Fig.\ \ref{Fig6}. This influences the unfolding procedure we use in Sect.\ \ref{resultsB}. However, for the most interesting cases such as the $({\rm a})$ and $({\rm f})$ this imprecision does not play a role.

With the increase of $m$ and for fixed $F_{\rm dc}$ (along the vertical dashed line in the right panel of Fig.\ \ref{Fig3}), the normalized LE spectrum undergoes a transition from a regime where all the exponents are negative, to the regime where half of the spectrum is positive. From Fig.\ \ref{Fig6} the normalized LE spectrum does not follow any characteristic function or rule other than slowly moving to the positive domain for increasing mass term. The particular island (dark orange patch in the heatmap) with roughly fifty percent of positive LEs where $\bar{\lambda} \rightarrow 0^{-}$ we observe a distribution that resembles a semi-circle law. This is also nicely illustrated in Fig.\ \ref{Fig5} ${\rm (b)}$ for the case $N = 50$. 

Similar findings, but in conservative models, have been reported in Ref.\ \cite{Gur-Ari_Hanada_Shenker_2016,Patra_Ghosh_2016} where the authors make the connection between the LE spectrum and RMT. They identify the semi-circle law which is symmetric around $\bar{\lambda} = 0$ and under proper rescaling matches the semi-circle law and that of GOE \cite{Livan_Novaes_Vivo_2018}. We argue that in a dissipative dynamical model such as the underdamped FK model the features of GOE are present even though we are not able to fit our results to the semi-circle due to dissipation that forces the spectrum to have a negative mean. In the following sections, we use diagnostic tools common in the RMT literature to do so.

\subsection{Normalized Lyapunov spectrum spacings} \label{resultsB}
Conventionally in the study of RMT and related problems, from the eigenvalue spectrum, the nearest-neighbor spacing distribution is generated. Rather than the eigenvalue spectrum density, the distribution of eigenvalue spacings is typically regarded to be universal \cite{Eynard_Kimura_Ribault_2018}. For our particular purposes, it is sufficient to present the following relevant probability density functions (PDFs)
\begin{center}
\begin{equation}
P_{\rm Poisson} (s) = e^{-s} \quad \qquad {\rm \textit{uncorrelated}}, \label{Poisson}
\end{equation}
\begin{equation}
P_{\rm GOE} (s) = \frac{\pi s}{2} e^{- \frac{\pi}{4} s^2}  \qquad {\rm \textit{correlated}}, \label{GOE}
\end{equation} 
\end{center}
where similar expressions exist for the remaining correlated Gaussian ensembles \cite{Dietz_Haake_1990}. Eq.\ (\ref{GOE}) in the RMT literature is recognized as the Wigner's surmise, while Eq.\ (\ref{Poisson}) is the well-known Poisson distribution.

LEs can be viewed as eigenvalues that characterize chaotic, periodic, and quasiperiodic motion \cite{Ahlers_Zillmer_Pikovsky_2001,Odavić_Mali_Tekić_2015}. Therefore, treating them as such the corresponding PDF can be computed. In our case, $P(s)$ represents the probability density of two consecutive LEs having spacing $s$, and as any PDF it is normalized to unity. Typically the spacings represent an order set of 
\begin{center}
\begin{equation}
 \tilde{s}_{i',j} = \lambda_{i' + 1,j} - \lambda_{i',j}, \label{def_space}
\end{equation}
\end{center}
for realization $j = 1, 2, ..., M$.  In practice, the LE spacings PDF is obtained from the normalized histogram of $s_{i',j}$ values as
\begin{equation}
s_{i',j} \equiv N \cdot \big( R(\lambda_{i' + 1,j}, \bar{\lambda}) - R(\lambda_{i',j},\bar{\lambda}) \big), \label{spacings}
\end{equation}
where $i' = 1, 2, ..., N - 1$ and with the help of the cumulative spacing distribution $R(\lambda,\bar{\lambda})$ defined as 
\begin{equation}
 \\ R(\lambda_{i',j},\bar{\lambda}) = \int\limits_{\bar{\lambda}}^{\lambda_{i',j}} \rho(\lambda') {\rm d} \lambda' . \label{cumu}
\end{equation}

\begin{figure}[] %\bigskip
	\includegraphics[width=\columnwidth]{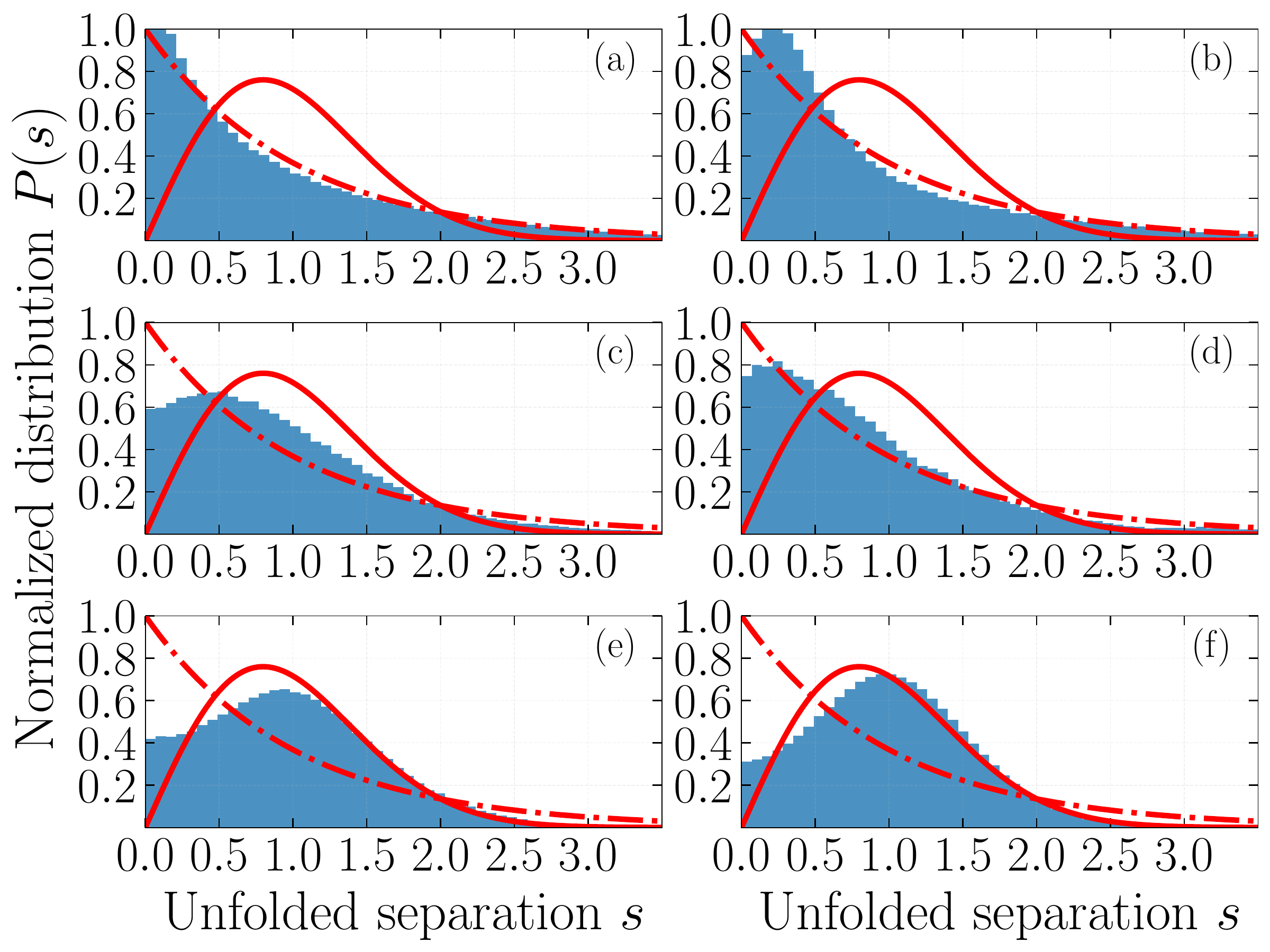}
\centering
\caption{\label{Fig7}(Color online) The unfolded LE spectrum histograms obtained using Eq.\ (\ref{spacings}) and (\ref{cumu}). Parameters used in simulations are the same as in Fig.\ \ref{Fig6}. The red dashed line is the Poisson distribution and full line is the Wigner's surmise, Eq.\ (\ref{Poisson}) and (\ref{GOE}), respectively. Further details are given in text.}
\end{figure}

In RMT literature this particular way of computing the spacings PDF is known as the ``unfolding'' procedure. The unfolded LE spectrum, computed using Eq.\ (\ref{spacings}), can now be directly compared to Eqs.\ (\ref{Poisson}) and (\ref{GOE}), and the results are presented in Fig.\ \ref{Fig7}. Note the difference between definitions of $s$ and $\tilde{s}$.  

In the strictly overdamped regime (see Fig.\ \ref{Fig7} $({\rm a})$), the unfolded spectrum histogram is consistent with the Poissonian distribution. Small fluctuations exist, but the LE spectrum spacings in $s \to \infty$ limit convincingly follow Eq.\ \ref{Poisson}. However, to achieve this identification a large number of particles ($N=400$) had to be used so the LE spectrum is sufficiently smooth. With further increase of the particle number, we expect the spacing spectrum to progressively move closer and closer to the Poisson distribution as it already showed this tendency for $N < 400$ and with the increase of $N$. We further comment on the physics of this regime in Sect.\ \ref{resultsC}.

In the underdamped regime and for $m = 0.5, 1.582, 2.3$ the unfolded separation distribution does not follow the Poisson law or the Wigner's surmise, see Fig.\ \ref{Fig7} ${\rm (b)},{\rm (c)}, {\rm (d)}$. For $m = 0.5$ the system exhibits similar periodic behavior as in the $m = 0$ case, and this is the reason why for $s > 2$ the separations between LEs appear to be Poissonian distributed. We checked that regions with similar chaoticity conform to similar distributions, i.e.\ regions with the same color in the heatmap in Fig.\ \ref{Fig3}, have equivalent LE spectrum and spacings. 
 
 \begin{figure*}[] %\bigskip
\includegraphics[width=18cm]{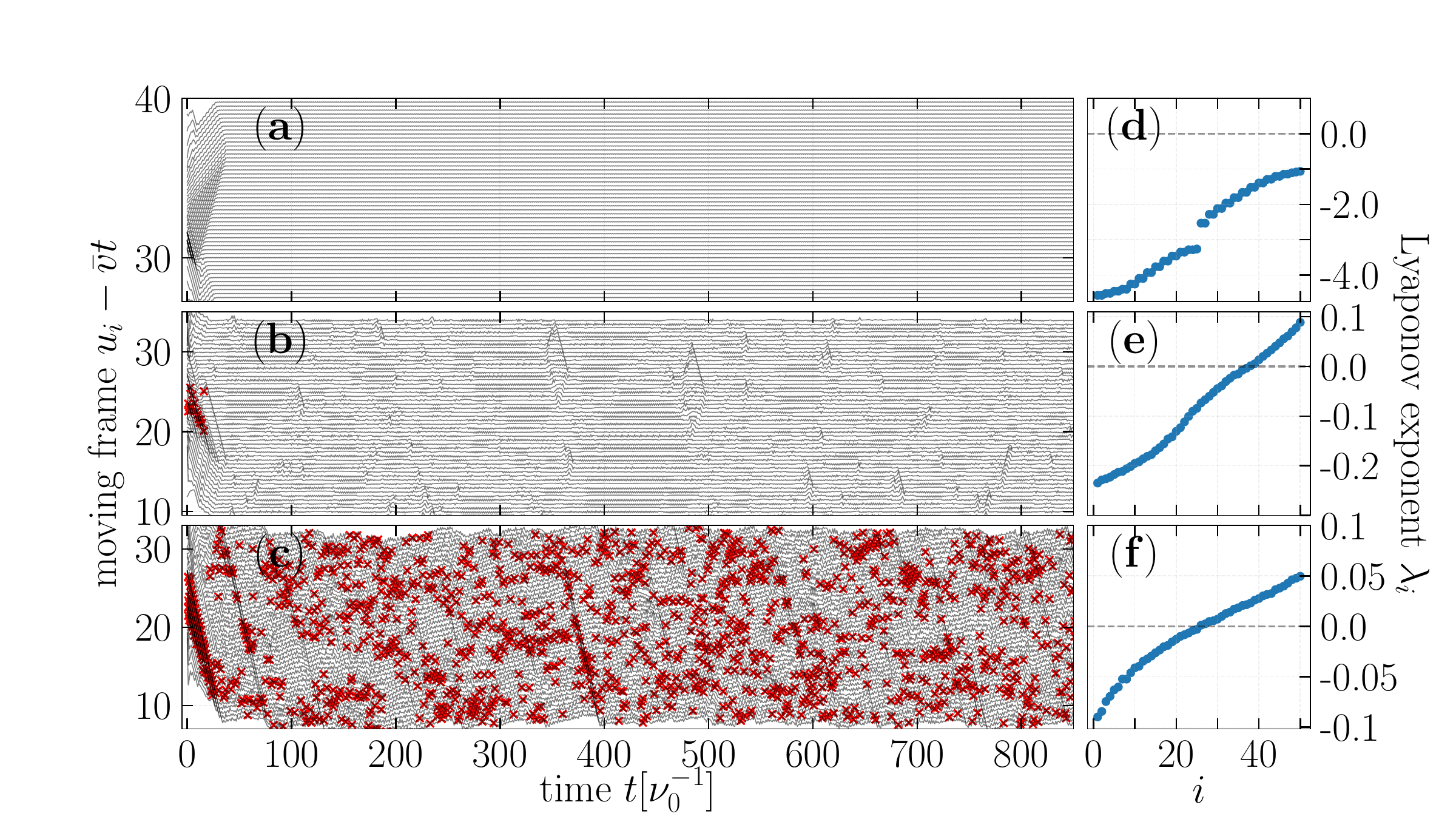}
\centering
\caption{\label{Fig8}(Color online) A single realization of particle trajectories for different mass parameters: $({\rm a})$ $m = 0$, ${\rm (b)}$ $m = 2.047$ and ${\rm (c)}$ $m = 4.653$ at fixed dc force $F_{\rm dc} = 0.27$ and remaining parameters as in Fig.\ \ref{Fig2}. The trajectories are plotted in the moving frame (sliding regime), for visual clarity. We used periodic boundary conditions as mentioned in Sect.\ \ref{model}. Red crosses specify when particles break the initial time ordering.   }
\end{figure*}

As the mass is increased the system enters the hyperchaotic regimes. In Fig.\ \ref{Fig7} ${\rm (e)}$ we observe a distribution depletion in the $0 < s < 1$ range which signifies that repulsion between the LEs occurs. LE repulsion was first observed in \cite{Ahlers_Zillmer_Pikovsky_2001}, in a conservative model which is used to describe the periodically kicked rotator. In the parameter regime where this model exhibits hyperchaos, the act of coupling between nearest-neighbor standard maps introduces LE repulsion. Before coupling, this model was already hyperchaotic but did not exhibit any LE level repulsion, hinting at the fact that coupling plays an extremely important role in generating a spectrum that exhibits \textit{correlated} RMT ensemble features. In the same paper, the authors mention that dissipative models also exhibit similar behavior, but in our work, we provide explicit evidence for this. In the dissipatively driven FK model, the coupling between particles is the nearest-neighbor and not treated as a parameter (see Eq.\ (\ref{model2})). Therefore, any hyperchaotic behavior within such a model is expected to exhibit LE repulsion. 

For $m = 4.75$ in Fig.\ \ref{Fig7} ${\rm (f)}$ the unfolded spectrum has an even stronger repulsion as $s \to 0$ compared to the other sampled cases. This parameter point sits in the already mentioned hyperchaotic regime with roughly $50 \%$ positive exponents (dark orange region in Fig.\ \ref{Fig3}). $P(s)$ in this case follows the Wigner surmise perfectly for $s > 2 $ indicating that this distribution is more similar to Eq.\ (\ref{GOE}) than (\ref{Poisson}). The maximal height of the distribution matches that of the GOE as well, but it is shifted to the right. We attribute the lack of a persuasive overlap to the dissipative dynamics that force the LE spectrum to have an asymmetric distribution. The phase space volume not being conserved through time evolution leads to such a shifted peak in the unfolded spectrum. We take this as an indication that in the dissipative system the LE spectrum might be described not by a pure GOE ensemble but rather by an ensemble mixture of GOE + Wishart ensemble \cite{Livan_Novaes_Vivo_2018}. An additional hint for this we can observe from Fig.\ \ref{Fig6} ${\rm (e)}$ and ${\rm (f)}$. In particular, the left tail of the normalized density $\rho (\lambda)$ shows similar behavior to the Marchenko-Pastur law \cite{Livan_Novaes_Vivo_2018,Mar_enko_1967}. 

The lack of strong LE repulsion as $s \! \to \! 0$ can not be exclusively attributed to the dissipative nature of the dynamics from our study. Inspecting the LE spectrum in more detail we often observe pairs of LEs to be present. These exponents are equivalent up to 2-3 decimal digits and persist to be so with the increase in the system size $N$. This kind of ``near degeneracies'' have also been observed in chains of R\"{o}ssler oscillators and are attributed to the (discrete) translational symmetry of the model \cite{PikovskyPoliti2016}. Similarly, we argue that due to this fact and the inherent fluctuating nature of finite-time LEs the Lyapunov vectors tend to align along with the same directions in the phase space producing the near degeneracy and ultimately the lack of complete LE repulsion. This is typically thought to be a spurious and subtle problem of numerical accuracy in the method for obtaining the LEs. For a definite and consistent analysis, the Lyapunov vectors have to be computed. This kind of analysis would also potentially  explain why in the conservative system of coupled Kuramoto oscillators, similar features are present, see Fig.\ 4.\ of \cite{Patra_Ghosh_2016}. Conversely, why and how these features are not present in the classical D0-brane model, see Fig.\ 1.\ \cite{Hanada_Shimada_Tezuka_2018}.

\subsection{Particle trajectories}
\label{resultsC}
In previous studies of the dissipatively driven FK model, the particle trajectories provided valuable information about the presence of chaos. Insight into spatiotemporal dynamics of the system is used, for example, in \cite{Strunz_Elmer_1998}, to supplement the common tool such as the LEs for chaos detection. 

In Fig.\ \ref{Fig8} $({\rm a})$,${\rm (b)}$, and ${\rm (c)}$ we plot the particle trajectories over time for different particle masses in the moving reference frame for visual clarity. Moving frame is used because the system is in the sliding regime where the particles exhibit collective motion related to ac+dc driving, see Fig.\ \ref{Fig4}. In the remaining figures of Fig.\ \ref{Fig8} ${\rm (d)}$, ${\rm (e)}$, and ${\rm (f)}$ the corresponding LE spectrum is presented. 

When the system is in the strictly overdamped regime ($m = 0$), i.e.\ when the inertial term is negligible in comparison to the damping one, individual particles act as impenetrable hard spheres. In Fig.\ \ref{Fig8} $({\rm a})$ this is depicted clearly, wherein the steady-state regime each particle feels the neighboring ones but they never exchange places and break the ordering at the initial time. Even in the transient regime, the particle order is preserved. To specify when two particles trajectories cross and particles break the initial time ordering we used red crosses, and no red crosses are shown in the strictly overdamped regime. This is because in this regime the physics of the model is subjected to Middleton's no-passing rule \cite{Middleton_1992}. 

Middleton's no-passing rule refers to order-preserving nature of the underlying dynamics in presence of convex interparticle interaction and is extensively exploited to gain analytical insight into complex dynamics \cite{FlorAP}.  Under such assumption, if two configurations are initially ordered $u = (u_{i})_{i \in \mathbb{N}}$ and $\tilde{u} = (\tilde{u}_{i})_{i \in \mathbb{N}}$, e.g.\ $u \leq \tilde{u}$ (where $\leq$ refers to each particle position $u_{i} \leq \tilde{u}_{i} \forall i$) then this ordering persists through time evolution \cite{Slijepčević_2015}. This means that the system does not evolve towards less complexity than it already possessed at the initial time. In  Ref.\ \cite{Odavic} it was argued that this rule is related to the absence of chaotic behavior. We now provide further evidence that this is indeed the case and additionally claim that it leads to Poissonian statistics in the LE spectrum. Explicit evidence can be found in Fig.\ \ref{Fig6} $({\rm a})$ and Fig.\ \ref{Fig7} $({\rm a})$ and Fig.\ \ref{Fig8}. 
 
Intuitively, in the strictly overdamped regime, the avoided crossings behavior of particle trajectories leaves less phase space for each of them to explore. This restricted phase space means that LEs are left with less variability and therefore the exponent values can and must be close together (but are not degenerate). This is nicely illustrated by the unfolded LE spectrum $P(s)$ from Fig.\ \ref{Fig7} ${\rm (a)}$ as $s \rightarrow 0$. In the underdamped regime, the available phase space is larger and LEs have inherently larger separations between each other, and thereby justifying the LE ``repulsion'', i.e.\ suppression of $P(s)$ weight as $s \rightarrow 0$, see e.g.\ Fig.\ \ref{Fig7} ${\rm (f)}$.

In the underdamped regime, due to the presence of the inertial term, the particles can overcome the harmonic interparticle repulsion and exchange positions. This is shown in Fig.\ \ref{Fig8} ${\rm (b)}$ and  ${\rm (c)}$ with the corresponding LE spectrums given in ${\rm (e)}$ and ${\rm (f)}$. Different chaotic regimes can be identified from Fig.\ \ref{Fig8} ${\rm (b)}$ compared to ${\rm (c)}$. First, one where the crossings happen only in the transient regime and the second one where such crossings can propagate through time and are present even in the steady-state regime. The latter is the regime with roughly fifty percent positive LEs (the dark orange region in Fig.\ \ref{Fig3}) and where the LE repulsion is prevalent. We note that particles after exchanging positions quickly revert to previous ordering due to the overwhelming strength of the particle interactions.  We checked that choosing different initial conditions, e.g.\ uniformly distributed particle positions, the transient dynamics looks different but in the steady-state, the general conclusions provided are still valid. This means that LE spectrum and the LE fluctuations are completely determined by the nonlinear dynamics rather than by the choice of initial conditions.

In literature, the term ``avoided crossing'' is typically used to describe the repulsive behavior of eigenvalues in quantum mechanical models with the change of some control parameter \cite{PikovskyPoliti2016}. Similar behavior, as already mentioned, has been used in the context of LEs \cite{Ahlers_Zillmer_Pikovsky_2001}. Note that the avoided crossing behavior discussed so far in this chapter is in terms of particle trajectories. We identify a dichotomy, where the lack of avoided crossing in the trajectories leads to avoided crossings in the LEs (exponent repulsion) and vice versa.  

Within the paradigm of the FK model, the particle trajectory crossing behavior might appear to be unphysical, as any classical object due to its finite size would forbid it. However, in this regime and under the following variable changes
$2 \pi u_{i} = \varphi_{i},  m = K = a^2,  m \alpha  = 1, mI = 2 \pi  F_{\rm dc}, m A = 2 \pi  F_{\rm ac}, \omega = 2 \pi \nu_{0}$ the dissipatively ac+dc driven FK model directly maps to the realistic model of Josephson transmission lines of stripline geometry \cite{Rahmonov_Tekić_Mali_Irie_Shukrinov_2020,eratum}. These types of models can capture quantum phenomena that occur in Josephson junction systems despite considered to be classical in essence \cite{OBBook,Pfeiffer_Abdumalikov_Schuster_Ustinov_2008}. In the JJ array models $\varphi_{i}$ represents the phase jump of the wave function across a single junction, and therefore in this particular context, there is no ambiguity in the crossing behavior as it simply implies that the phase order is violated.

\subsection{Consecutive Lyapunov exponent spacing ratio}
\label{resultsD}

\begin{figure}[] %\bigskip
\includegraphics[width=\columnwidth]{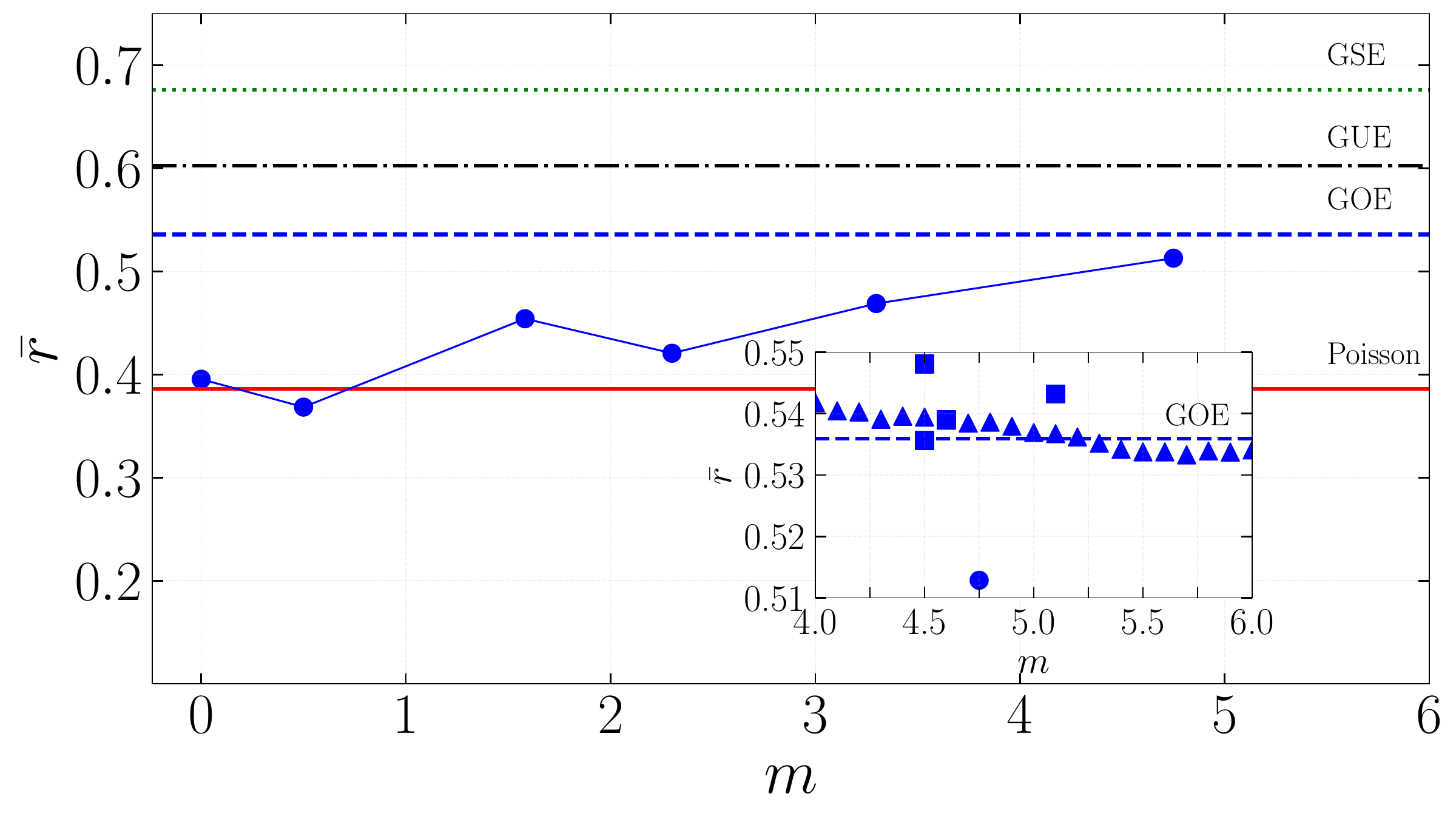}
\centering
\caption{\label{Fig9}(Color online) Consecutive LE spacing ratio $\bar{r}$ (blue circles) for the different mass parameters used in Fig.\ \ref{Fig6} and \ref{Fig7}. Lines between the values is a guide for the eye. The explicit expressions for the GUE and GSE ensembles are given in \cite{Atas_Bogomolny_Giraud_Roux_2013}, while the values for the Poisson and GOE ensembles are given in the text. The inset shows the $\bar{r}$ for different masses along the $F_{\rm dc} = 0.27$ line (triangles), and other randomly chosen parameter points in the hyperchaotic region with roughly $\sim  50 \%$ percent positive LEs (squares). The inset data has been generate for $N  =  50$ where $M  \sim  10^{4}$. The discussion of the inset and the figure is given in the text.}
\end{figure}

To supplement the standard tools that uncover the RMT statistics, such as the normalized density in Sect.\ \ref{resultsA} and the spectrum unfolding in Sect.\ \ref{resultsB}, we compute the average consecutive spacing ratio $\bar{r}$. The ratio is defined as
\begin{equation}
\bar{r} = \Bigg\langle \Bigg\langle \frac{{\rm min} (\tilde{s}_{i',j},\tilde{s}_{i'+1,j})}{{\rm max}(\tilde{s}_{i',j},\tilde{s}_{i'+1,j})} \Bigg\rangle \Bigg\rangle_{N-2,M},
\end{equation}
where $\tilde{s}_{i',j}$ is as defined in Eq.\ \ref{def_space}. First, the average is taken over $N - 2$ consecutive spacings values then additionally over the $M$ samples. The consecutive spacing ratio does not require unfolding and allows for a more transparent comparison between RMT ensembles and experimental/simulated datasets \cite{Atas_Bogomolny_Giraud_Roux_2013}. In the study of quantum many-body localization where an extremely large number of realization is computationally expensive to realize, the consecutive level spacing ratios PDF $P(r)$ was shown to provide more precise results than the common spacing distribution $P(s)$ \cite{Oganesyan_Huse_2007}. The average consecutive spacing ratio is independent of the local density of states and therefore a more suitable diagnostic tool than the unfolded spectrum in case the number of ensemble realization is limited. This happens, as mentioned, in the quantum many-body problems where sampling of the eigenspectrum is limited due to the exponentially increasing Hilbert space dimension, or in our particular case where we flow $\sim  N^2$ differential equations to obtain the LE spectrum of $N$ exponents.

In Fig.\ \ref{Fig9}, we present the results for the particular six points in the parameter space and statistics that were already the subject of investigation in Sects.\ \ref{resultsA} and \ref{resultsB}. In the strictly overdamped regime, the ratio $\bar{r}$ is consistent with the analytically predicted value $2 \ln{2} - 1$ corresponding to the Poisson distribution. In the underdamped regime and with the increase of the mass the consecutive LE spacings ratio slowly moves towards and close to $4 - 2 \sqrt{3}$ that corresponds to the GOE ensemble prediction \cite{Atas_Bogomolny_Giraud_Roux_2013}.

In the inset of Fig.\ \ref{Fig9}, we present the results of $\bar{r}$ for multiple points in the parameter region with roughly $\sim  50 \%$ percent positive LEs, i.e.\ parameter points in the dark orange region in Fig.\ \ref{Fig3}. The consecutive LE spacing ratio consistently shows ``agreement'' with the GOE ensemble. In this check, we used $N = 50$ to generate enough samples for a reliable evaluation. However, some points lie close or on top of the GOE expectation that after close examination of the unfolded spectrum does not fully show compliance with the Wigner's surmise expression. These results provide a false positive argument for a complete overlap with GOE, and should be used to state that the underlying LE spectrum is close to but surely not identical to GOE due to dissipation, finite-size effects, and insufficient statistics.  
\subsection{Distribution of the largest Lyapunov exponent}\label{resultsF}

Finally, we discuss the largest LE $\lambda_{\rm max}$ fluctuations in the hyperchaotic regime with roughly $50 \%$  positive exponents. In the common examples of chaotic dynamical systems, the distribution of individual finite-time LEs follows the Gaussian distribution \cite{Ott_2002} while, for example, in the presence of intermittent chaos this distribution was shown to be non-Gaussian \cite{Prasad_Ramaswamy_1999}. In the following, we argue that due to the presence of LE level repulsion (see Fig.\ \ref{Fig7}) the fluctuation statistics of the  $\lambda_{\rm max}$ show behavior similar to the Tracy-Widom (TW) distribution \cite{TracyWidom1994}. TW distribution, in effect, is a Gaussian distribution that is skewed and with particular left and right decaying tails. Before presenting our results of using sensitive numerical indicators towards the presence of the TW distribution we first briefly present its definition. The largest eigenvalue statistics of an $[N \times N]$ GOE approaches the TW CDF as
\begin{equation}
F (\lambda) = \lim\limits_{N \to \infty}  {\rm Prob} \Big( \big( \lambda_{\rm max} - \sqrt{2 N} \big) \big( \sqrt{2} \big) N^{1/6} \le \lambda \Big).
\end{equation}
For a transparent comparison between eigenvalue statistics of RMT and TW, see e.g.\  \cite{Edelman_Wang_2013}.

The fluctuation statistics of the largest/smallest eigenvalue of a random matrix around its mean follows the TW distribution and since this discovery, it was demonstrated that it plays an important role in many diverse fields such as mathematical physics, directed polymer physics, random growth models, finance, etc \cite{Majumdar_2007,Tracy_Widom_1994, Baik_Deift_Johansson_1999}. For example, in experiments of interface growth in thin films \cite{Takeuchi_Sano_Sasamoto_Spohn_2011} the interface height fluctuation statistics have shown to be in accordance with the solution of the Kardar-Parisi-Zhang (KPZ) equation which is precisely the TW distribution \cite{Sasamoto_Spohn_2010}. Depending on whether the growing interface is flat or curved, different universality classes such as GOE and GUE were observed to be present, respectively. In the context of RMT, the TW distribution successfully captures the crossover between two different phases present in the eigenspectrum. In particular, the eigenvalues of an RMT ensemble that follows the semi-circle law behave differently in the bulk compared to the edges of the spectrum. The left tail of the TW distribution describes the ``strong'' coupling regime of the bulk and the right tail the ``weak'' coupling regime present at the spectral edges \cite{Majumdar_Schehr_2014}. 

From Fig.\ \ref{Fig6} $({\rm f})$ the right spectral edge of LE spectrum in the regime with $50 \%$ positive exponents appears to have finite support and to be consistent with the semi-circle law. Moreover, several spatially extended dynamical systems with their respective Lyapunov exponents and vectors have been found to fall in the KPZ class \cite{PikovskyPoliti2016}. Therefore, naturally, we investigate if the distribution of $\lambda_{\rm max}$ in the dissipative and ac+dc driven FK model is distributed in accordance with the TW. In this respect, we do not settle with simply superimposing the fluctuation histogram of $\lambda_{\rm max}$ and TW PDF, but rather investigate this question using the following robust measures.

To reduce the statistical noise and to minimize small dataset sampling bias (present in any finite dataset such as our LE spectrum data) we average the statistical observables in the following way
\begin{equation}
\Big\langle A \big[ \lambda_{\rm max} (M') \big] \Big\rangle_{P} = \frac{1}{P} \sum\limits_{p = 1}^{P} A \big[ \lambda_{\rm max} (M'_{p}) \big] \label{averr}
\end{equation}
where $A$ denotes the observable of interest and $M'_{p}$ is the size of the $p$'th subset of the total dataset of size $M$. We sample $p = 1, 2, ..., P$ times $M' \le M$ values and finally average is taken. Performing averages with Eq.\ \ref{averr} is only beneficial in the case of small $M'$ and in case of large enough $M$ these averages should give the same output. 

\begin{figure}[] %\bigskip
\includegraphics[width=\columnwidth]{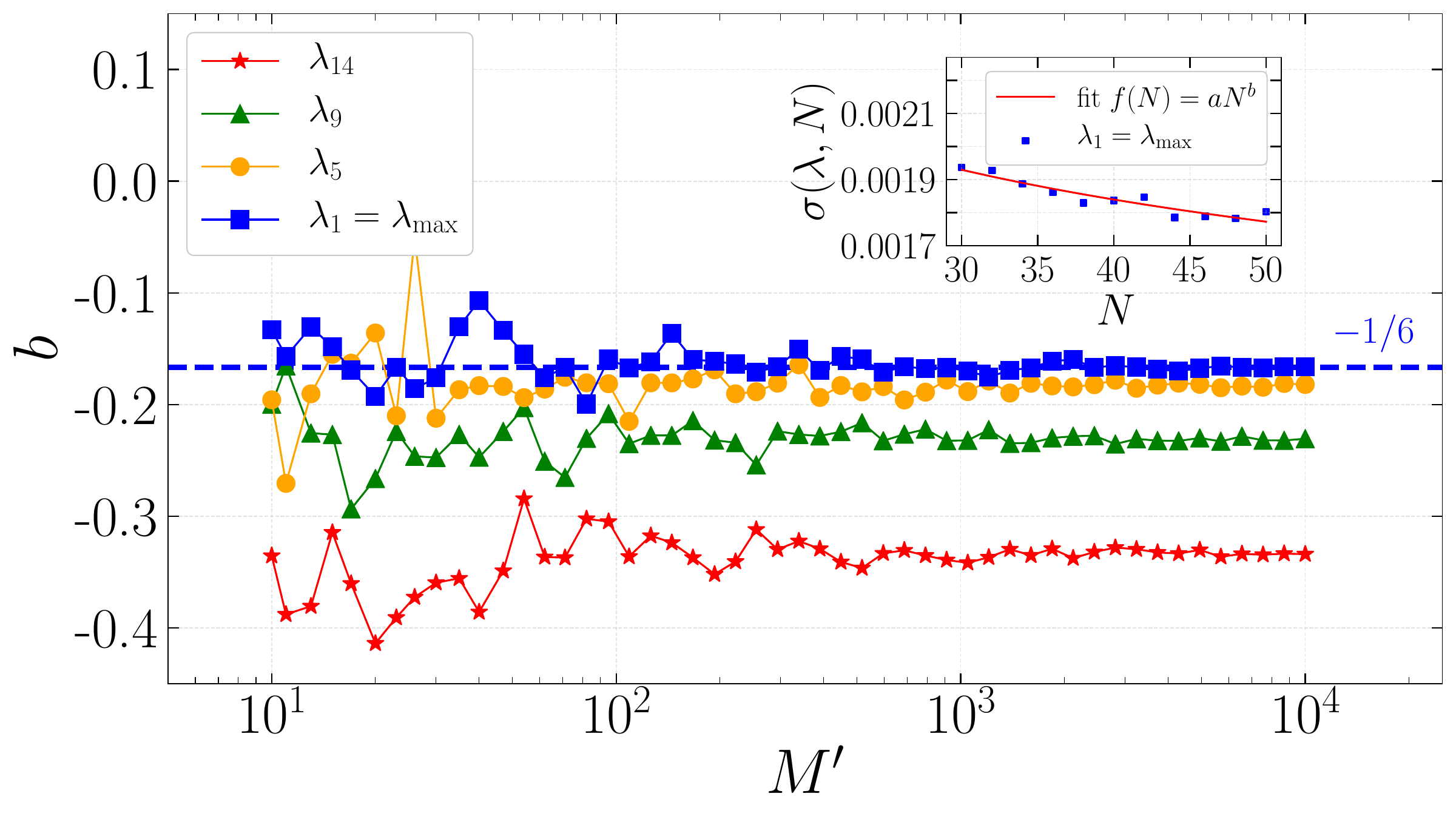}
\centering
\caption{\label{Fig10}
(Color online) Scaling exponent $b$ of the underlying fluctuation distribution for different ordered LEs  plotted against increasing number of simulation averaged realizations $M'$ with $P = 50$. To reduce the graphical clutter we selected to present LEs significantly far apart in the spectrum. With the straight dashed blue line the expectation from RMT spectral edge fluctuations of $b = -1/6$ is plotted for comparison. In the inset the standard deviation of the largest LE data is fitted for $M' = 10^{4}$ and changing particle number $N$ for the largest LE $\lambda_{1} = \lambda_{\rm max}$. Obtained fitting parameters are $a = 0.00339496(5)$ and $b = -0.1661(4)$. Remaining simulation parameters are: $F_{\rm dc} = 0.27$, $m = 4.653 $, $\omega = 1/2$, $F_{\rm ac} = 0.2$ and $\nu_{0} = 0.2$. Data points for the main frame are generate by performing the regression analysis as depicted in the inset.}
\end{figure}

In Fig.\ \ref{Fig10} we present the results for the standard deviation $\sigma$ of the distribution of the largest LE and a few other LEs in the bulk. We find that with sufficient sampling the $\lambda_{1} = \lambda_{\rm max}$ fluctuations in the spectral edge scale as expected from RMT (dashed blue line). In the inset we show that for the largest $M'$ the measured $\sigma$ in the LE spectrum follows the fitting function $f(N) \! = \! a N^{b}$, where $a$ is a trivial scaling parameter and $b$ is the power-law exponent. The observed scaling of the fluctuations as $ - 1/6$ is consistent with the TW distribution. Furthermore, LEs $\lambda_{k}, k \ge 1$ fluctuations away from the edge exhibit the same trend present in GOE bulk eigenvalues, i.e.\ moving toward $N^{-1/2}$ spectral bulk scaling.

It is important to note that, in generating the datasets required for the analysis in Fig.\ \ref{Fig10}, with the change of the particle number $N$ in the dissipative and driven FK model deep inside the underdamped regime the physics changes a bit as well. In particular, the average percentage of positive LEs changes with increasing the system size in this way. Therefore, a simple scan over different $N$  to obtain $\sigma$ invites caution. Fortunately, fixing the remaining model parameters (see caption of Fig.\ \ref{Fig10}) and for the selected $N$s the percentage fluctuation of positive LEs is less than $1 \%$.

 \begin{figure}[] %\bigskip
\includegraphics[width=\columnwidth]{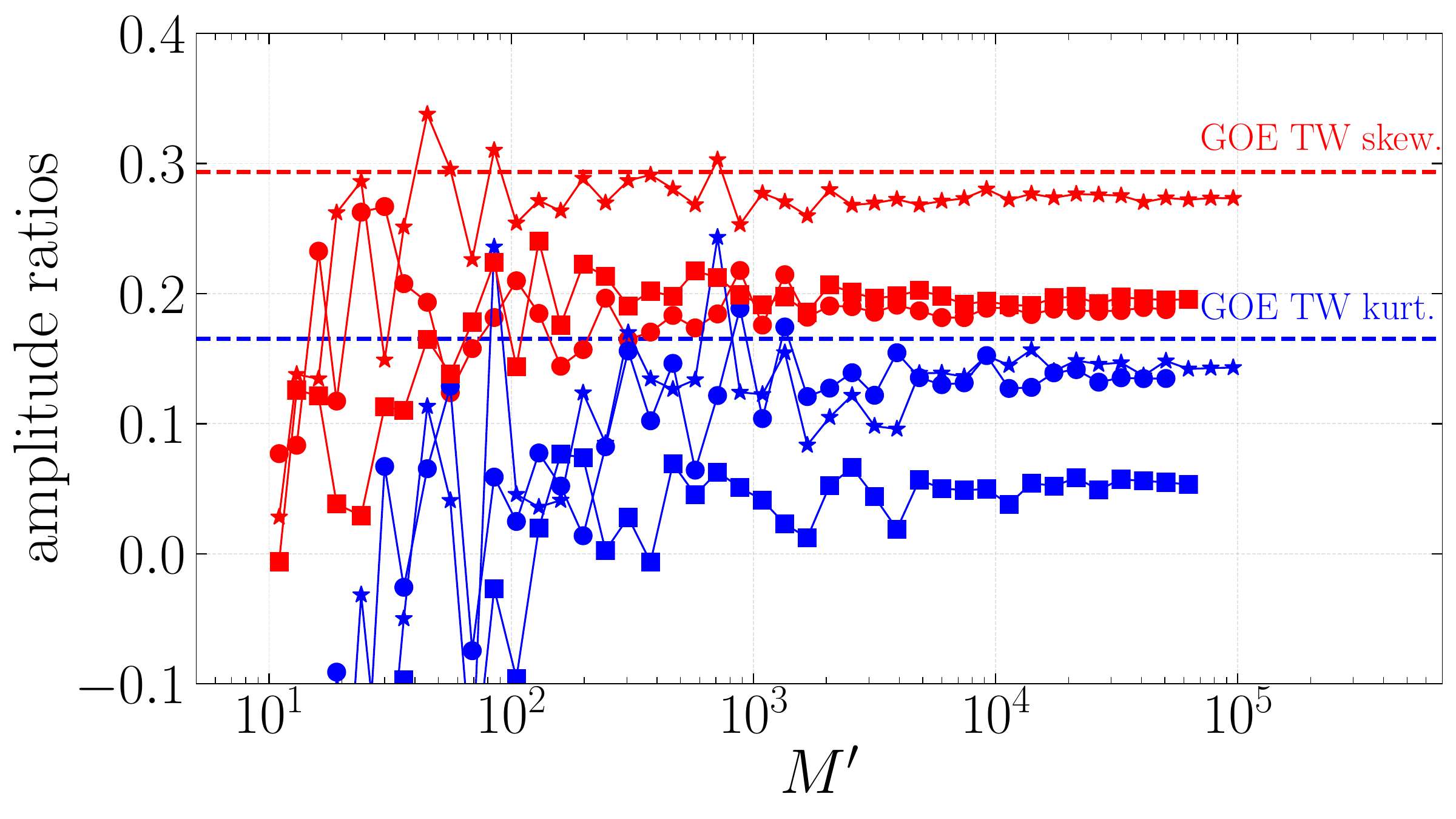}
\centering
\caption{\label{Fig11}(Color online) The skewness (red) and the excess kurtosis (blue) of the largest LE distribution for the $N = 50$ (circle) and $N = 100$ (square). We consider $m = 4.653$, $M = 53658$  for the $N = 50$ case and $m = 4.75$, $M = 66370$ for the $N = 100$ case, while the dc driving is fixed at $F_{\rm dc} = 0.27$ for which the systems is in the hyperchaotic regime with half of the LEs are positive. The remaining parameters are given in the caption of Fig.\ \ref{Fig6}. The straight dashed lines are the skewness and excess kurtosis of the GOE TW distribution.  For comparison we present the results of single $M = 10^{5}$ sampling of the largest eigenvalue of random GOE matrices (star) for $N = 100$. The number of subsequent averages takes is $P = 50$.}
\end{figure}

In Fig.\ \ref{Fig11} we compute the skewness and excess kurtosis which  measure the asymmetry of a PDF about its mean and its ``tailedness'', respectively. These kinds of measures allow for a test towards the presents of the TW distribution, e.g.\ see \cite{Sasamoto_Spohn_2010}. We employ Pearson's second skewness coefficient and the coefficient of excess kurtosis defined as 
\begin{equation}
{\rm Skew} = \frac{3 \big( \langle \lambda_{\rm max} \rangle_{{M'}} - \Lambda_{\rm max} \big) }{\sigma}, \label{skew}
\end{equation}
\begin{equation}
{\rm Kurt} = \frac{\frac{1}{M'} \sum\limits_{j = 1}^{M'} \big( \lambda_{{\rm max},j} - \langle \lambda_{\rm max} \rangle_{M'} \big)^{4} }{ \Bigg[ \frac{1}{M'} \sum\limits_{j = 1}^{M'} \big( \lambda_{{\rm max},j} - \langle  \lambda_{\rm max} \rangle_{M'}\big)^{2} \Bigg]^{2} } - 3 \label{kurt},
\end{equation}
where $\sigma$ defines the standard deviation and $\Lambda_{\rm max}$ is the sample median \cite{Kokoska}. The observed skewness and excess kurtosis for the distribution of the largest LE are not in complete agreement with TW expected values. Nevertheless, both measures have a statistically significant trend of having positive values. Furthermore, the skewness shows a consistent tendency of moving closer to the TW skewness with an increase of $N$, while this is not the case for the excess kurtosis. This is because excess kurtosis quantifies the behavior of the decaying tails of a given distribution and can only be improved by sampling the rare events which in turn requires numerical effort at a significantly larger scale compared to the one we were able to perform. We note that the characteristic skewness and the shape of the tails of the largest GOE eigenvalue distribution (which is known to follow the asymptotic TW distribution) are numerically difficult to obtain using finite $N$ and straight-forward sampling. This can be seen in Fig.\ \ref{Fig11} where the star-shaped data points represent such an attempt. In this case, the correct asymptotic skewness and excess kurtosis have been extracted in Ref. \cite{Borne} using specialized techniques.

%==================================================================
\section{Conclusion}\label{conclusion}
By studying the Lyapunov exponent fluctuations, and treating the exponents as eigenvalues that characterize the nature of the classical dynamics (chaotic, quasi-periodic, and periodic), in this paper, we provided numerical evidence for the existence of level repulsion akin to the RMT GOE ensemble in the ac+dc dissipatively driven Frenkel-Kontorova  model. The normalized Lyapunov exponent spectrum, Lyapunov exponent spacings, and consecutive Lyapunov exponent spacing ratio consistently indicate the presence of both \textit{uncorrelated} and \textit{correlated} statistics features in this model. Furthermore, the extensive numerical study of large parameter space revealed rich and varied dynamical phases. In the strictly overdamped regime (when particle masses $m \! \to \! 0$) the particle trajectories show avoided crossing behavior, whereas deep inside the underdamped regime and for a particular parameter set we identified persistent trajectory crossing behavior. We show that these two regimes have Poisson and almost Wigner GOE distributed Lyapunov exponent spacings, respectively. 

Our research is a step forwards to potentially explaining why classical dynamical models, such as Josephson junction array models subjected to external radiation (which represent related systems to the dissipatively driven Frenkel-Kontorova model), can capture quantum phenomena so well \cite{OBBook,Ustinov,Pfeiffer_Abdumalikov_Schuster_Ustinov_2008,Rahmonov_Tekić_Mali_Irie_Shukrinov_2020}. In particular, the underlying statistics (Lyapunov exponents in classical, and eigenvalues in the quantum case) seem to be guided by the same universal features. This opens a new and interesting direction in this and related fields of research.

Furthermore, the identification of Poisson statistics in the particular limit of the strictly overdamped regime of the Frenkel-Kontorova model gives hints to the potential presence of integrable structures in the continuum limit of the damped and driven perturbed sine-Gordon equation \cite{McLaughlin_Scott_1978}. However, the particular role of the Lyapunov exponents and their relation to integrability requires further investigations as was done in e.g.\ \cite{Poilblanc_Ziman_Bellissard_Mila_Montambaux_1993} for eigenvalues spacings in the quantum models.
We conjecture that the presence of uncorrelated statistics to be a general feature which means that \textit{in any strongly coupled overdamped system (with a sufficiently large number of degrees of freedom) where Middleton's no-passing rule applies the spectrum of LE spacings has Poisson statistics}. Moreover, in quantum many-body problems the presence of Poissonian statistics implies the existence of an infinite number of conservation laws which typically leads to an exact solution by using the Bethe ansatz technique. Therefore it would be interesting to explore the contingency of our results to the existence of integrability in the quantized version of the strictly overdamped version FK model if such a model can consistently be constructed. 

An interesting aspect of the hyperchaotic regime with $\sim 50  \%$ positive LEs is the presence of breathers. Looking at the trajectories in more detail reveals the presence of short-lived collective (that involves more than one particle) excitations that appear to mediate the particle crossings. From Fig.\ \ref{Fig8} ${\rm (f)}$ this dynamical mechanism is not obvious and  a more careful study, as in \cite{Laffargue_Lam_Kurchan_Tailleur_2013}, of this interesting phenomenon and its relation to LEs is required.

The main challenge we faced in this line of research was obtaining sufficient statistics on the Lyapunov spectrum fluctuations for a reliable comparison with asymptotic results of the Gaussian random ensembles. In the manuscript, we extensively discuss this fact and show how, for the examined model, we have to flow $\sim  N^2$ differential equations in the strongly nonlinear and chaotic regime ($N$ is the number of particles) to determine the spectrum. Interestingly, thanks to the sophisticated diagnostic tools developed in the literature, even with finite $N$ and a modest number of realizations $M$ we were able to capture the main features which are extremely close to the prediction of the Gaussian random matrix ensembles in the limits $N  \to  \infty$ and $M  \to  \infty$. Moreover, we presented hints that the largest Lyapunov exponent fluctuations, for a particular parameter regime, behaves according to the famous Tracy-Widom distribution.  Therefore, an imperative future research direction is the application of efficient algorithms to investigate if this indeed the case and how different degrees of hyperchaoticity present in the model influence the behavior of the largest Lyapunov exponent.

\label{concl}

%==================================================================================================================
\section*{Acknowledgments}
Jovan Odavi\' c would like to express his gratitude to the organizers of \textit{SFT-Paris-2019: Lectures on Statistical and Condensed Matter Field Theory} school and the hospitality at the Institute Henri Poincar\'{e} where interesting discussions on the subject of this work took place. Furthermore, we would like to thank Jorge Kurchan, Toma\v{z} Prosen, and Fabio Franchini for fruitful discussion. The authors gratefully acknowledge the AXIOM HPC facility and support provided by the Scientific Computing Research Group (SCORG) \cite{Skrbic} at Faculty of Sciences, University of Novi Sad. 

This work was supported by the Ministry of Education, Science and Technological Development of the Republic of Serbia under the grant 451-03-68/2020-14/200125, and the Croatian Science Foundation under the grant HRZZ IP-2016-06-1142.

\begin{appendix}
\section{Discussion on the total Lyapunov  exponent spectrum}\label{appendixA}

\begin{figure}[] %\bigskip
\includegraphics[width=\columnwidth]{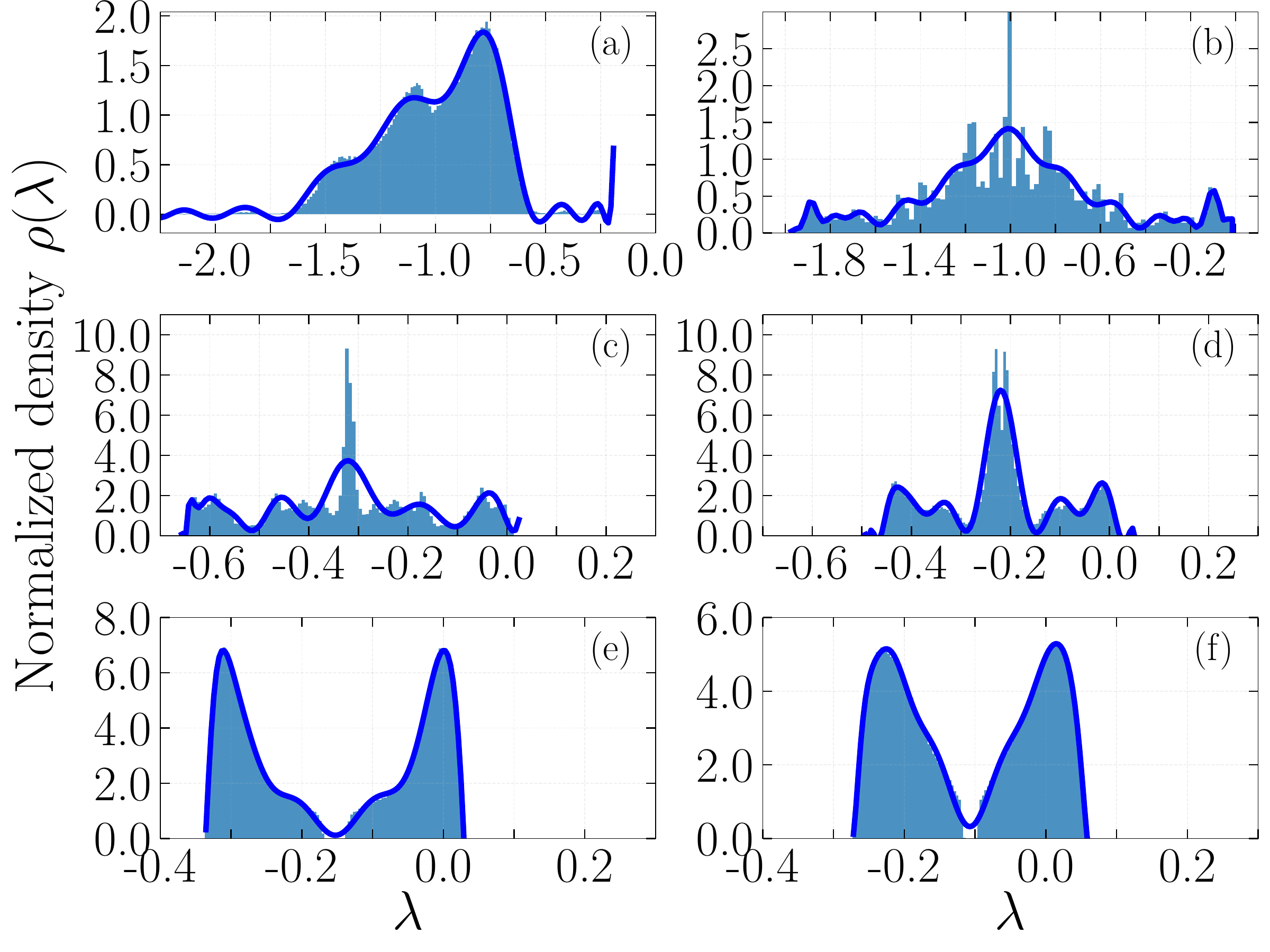}
\centering
\caption{(Color online) Normalized total LE spectrum density of $2N$ exponents at $F_{\rm dc} = 0.27$, $\omega = 1/2$, $F_{\rm ac} = 0.2$ and $\nu_{0} = 0.2$ for different masses: $({\rm a})$ This figure is the same as in Fig.\ \ref{Fig6} (a) $m = 0$, $N = 400$ and $M = 790 $, ${\rm (b)}$ $m = 0.5$, $N = 100$ and $M = 2116$, ${\rm (c)}$ $m = 1.582$, $N = 100$ and $M = 3433$, ${\rm (d)}$ $m = 2.3$, $N = 100$ and $M = 3309$, ${\rm (e)}$ $m = 3.298$, $N = 100$ and $M = 2815$ and finally ${\rm (f)}$ $m = 4.75$, $N = 100$ and $M = 2825$. The corresponding averages are: $({\rm a})$ $\bar{\lambda} \simeq  -0.995$, ${\rm (b)}$ $\bar{\lambda} \simeq -0.988$, ${\rm (c)}$ $\bar{\lambda} \simeq -0.316$, ${\rm (d)}$ $\bar{\lambda} \simeq -0.217$, ${\rm (e)}$ $\bar{\lambda} \simeq -0.151$ and ${\rm (f)}$ $\bar{\lambda} \simeq -0.105$. The dark blue lines are the fits obtained via Eq.\ (\ref{fit}) and represent the PDFs of the respected spectrum. Number of histogram bins is 100. \label{Fig12}  }
\end{figure}

In this Appendix, we discuss the total Lyapunov exponent spectrum measured for the considered variant of the Frenkel-Kontorova model in the underdamped regime. More specifically, we present the results for the full spectrum of $2N$ exponents instead of $N$ largest exponents used in the manuscript. 

From Fig.\ \ref{Fig12} we observe that the spectrum of the remaining $N$ exponents (not displayed in Fig.\ \ref{Fig6}) presents a mirror image of the spectrum of the largest $N$ exponents implying equivalent exponent spacings statistics. We note that considering only the first $N$ largest exponents for a large enough ensemble already provides sufficient data for our claims. For more details on the notations and parameters mentioned in the captions of Fig. \ref{Fig12}, please see Sect. \ref{results}.

\end{appendix}

%==================================================================================================================

%===========================

\section*{References}
\bibliographystyle{iopart-num}
\bibliography{Refs.bib}

\end{document}